\documentclass[11pt]{article}
\usepackage{fullpage, graphicx,amsmath,amssymb,cancel,float,color,mathrsfs,amsfonts,eufrak,amsthm}
\usepackage{bbm}
\usepackage{subfigure}
\usepackage{cite}
\usepackage{enumerate}
\usepackage{blindtext}
\usepackage{abstract}

\newtheoremstyle{remboldstyle}
  {}{}{}{}{\bfseries}{.}{.5em}{{\thmname{#1 }}{\thmnumber{#2}}{\thmnote{ (#3)}}}
\theoremstyle{remboldstyle}

\newcommand{\drop}[1]{}

\begin{document}
\title{Simultaneous Grain Boundary Motion, Grain Rotation, and Sliding in a Tricrystal}
\author{Anup Basak and Anurag Gupta\thanks{ag@iitk.ac.in (corresponding author)}}
\date{{\small Department of Mechanical Engineering, Indian Institute of Technology,
Kanpur 208016, India
\\  \today}}

 \maketitle

\begin{abstract}
Grain rotation and grain boundary (GB) sliding are two important mechanisms 
for grain coarsening and plastic deformation 
in nanocrystalline materials. They are in general coupled with GB migration and the resulting dynamics, driven by capillary and external stress, is significantly affected by the presence of junctions. Our aim is to develop and apply a novel continuum theory of incoherent interfaces with junctions to derive the kinetic relations for 
the coupled motion in a tricrystalline arrangement. The considered tricrystal consists of a columnar 
grain embedded at the 
center of a non-planar GB of a much larger bicrystal made of two rectangular grains. 
We examine the shape evolution of the embedded grain numerically using a finite 
difference scheme while emphasizing the role of coupled motion as well 
as junction mobility and external stress. The shape accommodation at the GB, 
necessary to maintain coherency, is achieved by allowing for GB diffusion along the boundary.

\vspace{4mm}\noindent {\bf Keywords:} Coupled grain boundary motion; Grain rotation; Grain boundary sliding;
 Triple junction; Tricrystal; Nanocrystalline material
\end{abstract}

\section{Introduction}
\label{intro_junction_2D}
\begin{figure}[t!]
\centering
\subfigure[]{
  \includegraphics[width=2.1in, height=1.6in] {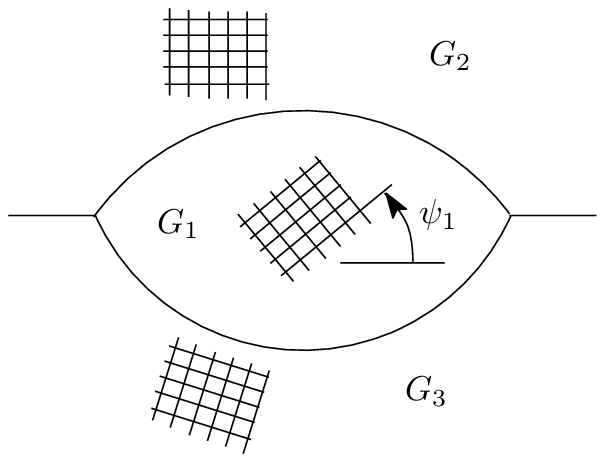}
	\label{schematic_rotation_grain1}}
	\hspace{10mm}
    \subfigure[]{
    \includegraphics[width=2.1in, height=1.6in] {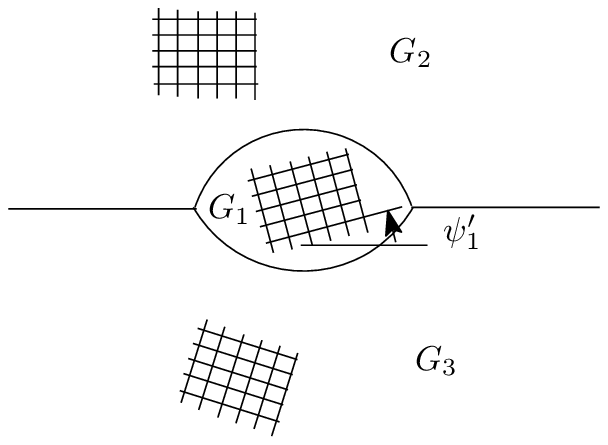}
    \label{schematic_rotation_grain2}}
\caption{A Schematic to depict the coupling between GB motion and rotation of grain $G_1$ under
GB capillary force in 
a tricrystal. Diagram (a) shows the initial configuration which evolves to (b) at a later time. 
The outer grains $G_2$ and $G_3$, being much larger than $G_1$, are taken to be
stationary (after \cite{trautt3}).}
\label{schematic_rotation_grain}
\end{figure}

Grain boundaries (GBs) and junctions play an important role in various deformation processes within  nanocrystalline 
(NC) materials which have an average grain size of few tens of nanometers and hence contain a large volume fraction of boundaries and junctions. The microstructural evolution in NC materials, especially during grain coarsening and plastic deformation, is dominated by grain rotation and relative grain translation coupled with GB migration \cite{meyers1,wang1,koch1, harris1}. The resulting motion is called coupled GB motion \cite{cahn1,taylor1}. The presence of triple junctions, which can occupy up to $3\%$ volume fraction in NC materials when the average grain size is around $10$ nm (Chapter 5 of \cite{koch1}), induces drag on GB migration and 
affects the coupled motion in a significant way \cite{czubayko1, wu1}. For an illustration of the coupled motion consider an isolated 
tricrystal arrangement, as shown in Figure \ref{schematic_rotation_grain1},
where a grain is embedded at the center of the planar GB of a large bicrystal. In the absence of external
stress, the embedded grain spontaneously rotates due to GB capillarity, thus changing its 
orientation, while shrinking to a size shown in Figure \ref{schematic_rotation_grain2}. 
The embedded grain can disappear either by shrinking to a vanishing volume or by reorienting 
itself to one of the neighboring grains. Grain rotation can be accomplished through
either a pure viscous sliding, or a tangential motion geometrically coupled with GB migration, or a 
combination of both \cite{cahn1,taylor1}. If the tricrystal is subjected to external stress, 
the grains can accomplish relative translational motion as well \cite{trautt3}; the center of
rotation of the embedded grain then need not remain fixed in space.

Our aim is to develop a thermodynamically consistent framework to study the coupled GB motion 
in the presence of triple junctions as driven by GB capillarity and 
external stress. More precisely, the main results of the present contribution are:

\noindent i) Developing a novel continuum framework, restricted to two dimensions, to study the dynamics of incoherent interfaces with junctions. An irreversible thermodynamical theory of incoherent interfaces, excluding junctions, has been previously developed by Cermelli and Gurtin \cite{cermelli2}. On the other hand, junctions have been studied only with respect to coherent interfaces \cite{simha1}. Furthermore, these previous studies were based on the configurational mechanics framework which requires \textit{a priori} postulation of configurational forces and their balances. In the present formulation the configurational forces appear as mechanisms of internal power generation so as to ensure that the excess entropy production remains restricted to only interfaces and junctions.

\noindent ii) Extending the existing theory of coupled GB motion to include triple junctions and relative tangential translation. The earlier work on coupled motion was restricted to bicrystals with a grain embedded within a larger grain such that the center of rotation of the embedded grain remains fixed
\cite{cahn1,taylor1,basak1}. The possibility of including junctions and relative translation of the grains was ignored in these models. These extensions were nevertheless mentioned by Taylor and Cahn \cite{taylor1} in their list of open problems related to coupled GB motion.  

\noindent iii) Performing numerical simulations for shape evolution of grains and GBs during coupled motion. Towards this end, we consider a tricrystalline arrangement ( as shown in Figure \ref{tricrystal_2D_stress}) and solve the coupled kinetic relations
for GB motion, rotational and translational movements of the grains, and junction dynamics. The dynamical equations are solved using a finite difference scheme adapted from a recent work on triple junctions of purely migrating GBs \cite{fischer1}. Our results are qualitatively in agreement with a recent paper \cite{trautt3} concerned with molecular dynamics (MD) simulations of the coupled motion in a tricrystal. 

We assume that the grains are rigid, free of defects, and do not posses any stored energy. 
The defect content as well as the energy density are confined to grain boundaries. 
Isothermal condition is maintained throughout. The assumption of grain rigidity is justified 
 since we consider the magnitude of the external stress to be much lower than the yield stress.
Also, in the present scenario the GBs do not exert any far-field stress and GB capillary 
exerts very low pressure on the neighbouring grains. The shape accommodation process, 
required to avoid nucleation of void or interpenetration of the grains at the GBs during relative 
rotation of the embedded grain, 
is controlled by allowing for diffusion along the GBs. Bulk diffusion in the grains, as well 
as across the GBs, is taken to be negligible compared to GB diffusion \cite{koch1}. Furthermore,
since both GBs and grains move at much smaller velocities
than the velocity of sound in that material, the inertial effects are ignored. 
The above assumptions provide the simplest setting to pursue a rigorous study of coupled 
GB dynamics. 

The paper has been organized as follows.
After developing the pertinent thermodynamic formalism in Section \ref{thermo_junction_2D}, the 
relevant kinetic relations for the tricrystalline configuration are derived in 
Section \ref{kineticsss_tricrystal_stress}.  The numerical results
are presented in Section \ref{results_junction_2D}. Finally, Section 
\ref{conclusion_junction_2D} concludes our communication.

\section{Thermodynamic formalism}
\label{thermo_junction_2D}
The dissipation inequalities at GBs and junctions are now derived within the 
framework of Gibbs thermodynamics, where various thermodynamic quantities (such as energy, 
entropy, etc.) defined over interfaces and junctions are understood as excess quantities of 
the system.  We begin by fixing the notation before deriving the consequences of the second law of thermodynamics in terms of various 
dissipation inequalities.

\begin{figure}[t!]
\centering
\subfigure[]{
  \includegraphics[width=3in, height=1.8in] {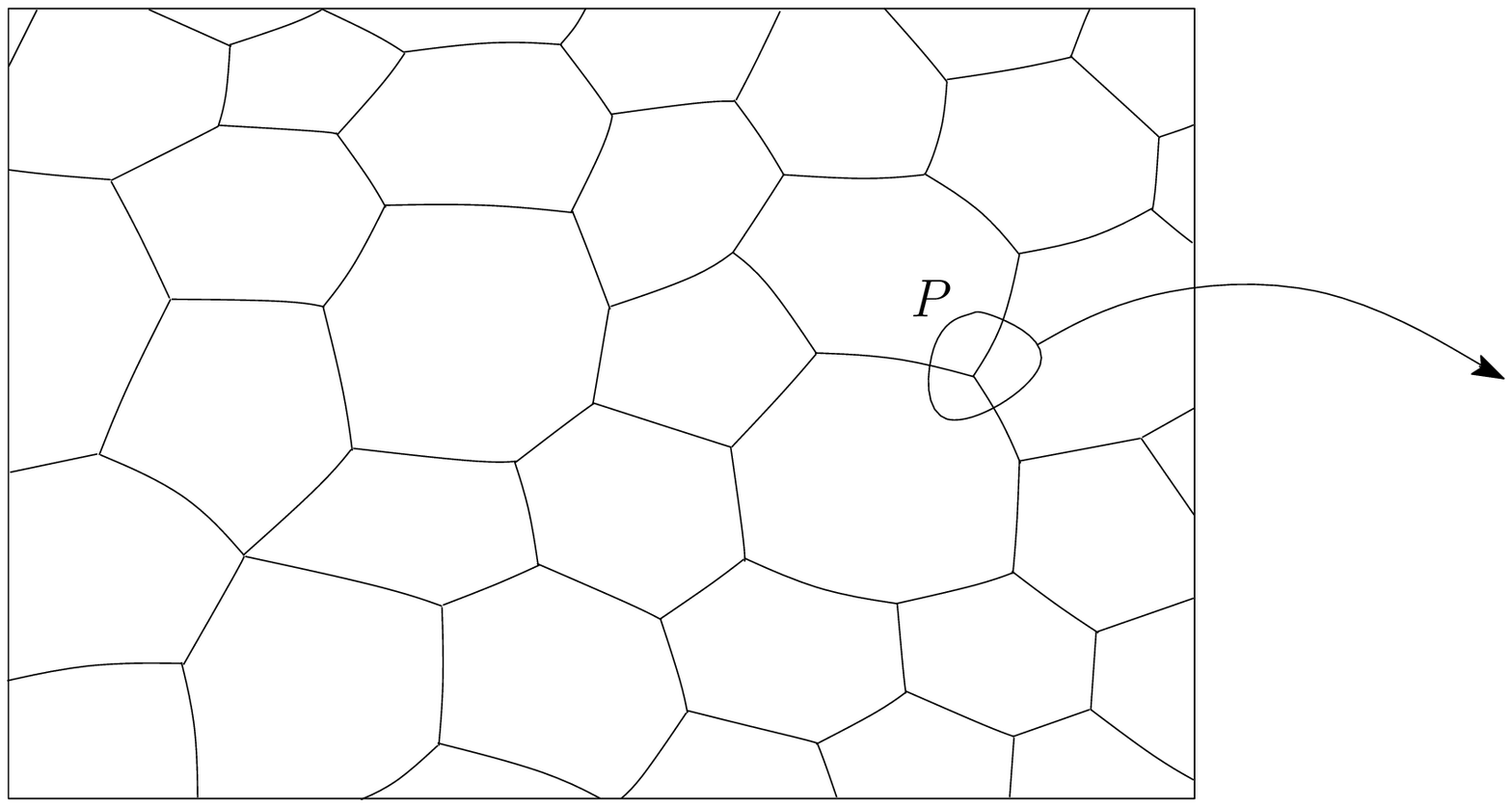}
	\label{polycrystal}}
	\hspace{10mm}
    \subfigure[]{
    \includegraphics[width=2.4in, height=1.9in] {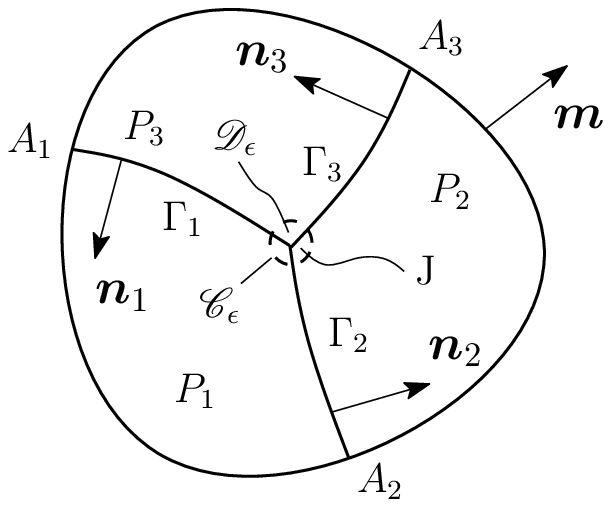}
    \label{control_vol_junc_2D_stress}}
\caption{(a) Schematic of a polycrystal with a region $P$. (b) The region $P$
containing three subgrains $P_i\,\, (i=1,2,3)$, three GBs $\Gamma_i$, 
and a junction $J$. Normal vectors for the outer boundary of  $\partial P$ 
and $\Gamma_i$ are denoted by ${\boldsymbol m}$ and ${\boldsymbol n}_i$, respectively. 
Point $A_i$ denote the edge of $\Gamma_i$ lying on $\partial P$. The broken circle 
${\mathscr C}_\epsilon$ is the boundary of the circular disc ${\mathscr D}_\epsilon$ which is
excluded to obtain the punctured domain $P_\epsilon=P\backslash {\mathscr D}_\epsilon$. }
\label{control_vol_junc}
\end{figure}

Consider a 2D region $P$ as shown in Figure \ref{control_vol_junc_2D_stress}
containing three domains $P_1$, $P_2$ and $P_3$, and a junction $J$. $P$ can be thought of as a 
subdomain in a polycrystalline material, as depicted in Figure \ref{polycrystal}. The boundary 
separating ${P}_i$ and ${P}_{i-1}$ $(i=1,2,3)$ has been represented by $\Gamma_i$ ($P_0$ is identified
with $P_3$). The normal ${\boldsymbol n}_i$ to $\Gamma_i$ is chosen such that it points into ${P}_i$.
The outer boundary of $P$ and the associated outward normal are denoted by $\partial P$ and 
${\boldsymbol m}$, respectively. We parametrize each of the GBs $\Gamma_i$ by an arc-length 
parameter $s_i$ which initiates at $J$ and increases towards the edge $A_i$. The tangent 
${\boldsymbol t}_i$ to $\Gamma_i$ is aligned in the direction of increasing $s_i$. The stress 
field at the junction is usually singular (cf. \cite{simha1} and Part H of 
\cite{gurtin_config}) and therefore all the analysis is restricted to a punctured domain 
$P_\epsilon$ which is obtained by excluding a small circular disc ${\mathscr D}_\epsilon$ 
of radius $\epsilon$ centered at the junction, i.e. $P_\epsilon=P\backslash{\mathscr D}_\epsilon$. 
We denote the periphery of the circular hole in $P_\epsilon$ by ${\mathscr C}_\epsilon$ 
whose normal ${\boldsymbol m}$ directs inside $P_\epsilon$. 
The velocity of the curve ${\mathscr C}_\epsilon$ approaches the velocity of the junction, 
denoted by ${\boldsymbol q}$, in the limit $\epsilon\to 0$. 

Let $f$ be a field defined in $P$ such that it is continuous everywhere except across
$\Gamma_i$. The 
jump in $f$ across $\Gamma_i$ is denoted by $[\![f]\!]=f^+-f^-$, where $f^+$ is the 
limiting value of $f$ as it approaches 
$\Gamma_i$ from the side into which ${\boldsymbol n}_i$ points and $f^-$ otherwise. 
The normal time derivative of a field $g$ defined on $\Gamma_i$ is given by 
\cite{gupta1}
\begin{equation}
\mathring{g} = \dot{g}+ V_i\,\nabla g\cdot{\boldsymbol n}_i
 \label{normal_time_derivative3}
\end{equation}
(no summation for the repeated index `$i$' is considered 
here and thereafter),
where the superposed dot stands for the material time derivative, $V_i$ is the normal velocity 
of $\Gamma_i$, and $\nabla$ is the gradient operator. 
 It represents the time rate of change of $g$ with 
respect to an observer sitting on $\Gamma_i$ and moving with the interface in its normal 
direction. 

\paragraph{Dissipation inequality}
We now derive the dissipation inequalities for the grains, GBs, and junction using the balance relations for  
mass and linear momentum given in Appendix \ref{balance_laws_stress_2D1}. 
In confirmation with the second law of thermodynamics for isothermal processes, the rate of 
change of free energy of the GBs is less than or equal to the total power supplied to $P$:
 \begin{equation}
\sum_{i=1}^3\frac{d}{dt}\int_{\Gamma_i}\gamma_i\,dl\leq \int_{\partial P}{\boldsymbol\sigma}
{\boldsymbol m}\cdot{\boldsymbol v}\,dl
-\sum_{i=1}^3(\mu\,{h}_i)_{A_i}+\sum_{i=1}^3({\boldsymbol c}_i\cdot{\boldsymbol w}_i)_{A_i},
\label{dissi_ineq_junc1}
\end{equation}
where  $\gamma_i$ is the energy density of GB $\Gamma_i$, 
${\boldsymbol\sigma}$ is the symmetric Cauchy stress in the grains, $\mu$ is the chemical potential of 
the atoms, $h_i$ is the diffusion flux along $\Gamma_i$, and $dl$ is an infinitesimal length 
along the boundaries. As noted before, we have ignored bulk free energy as well as volumetric diffusion 
in the grains. The vector ${\boldsymbol w}_i$ 
stands for the velocity of edge $A_i$ and ${\boldsymbol c}_i$ is the force conjugate associated
with it. The nature of the latter is elaborated below. The first term on R.H.S. of the 
inequality is the power input into $P$ due to external stress field on its boundary. The 
second term represents the power input due to additional mass flow. The third term, which is
non-standard, represents the power input into $P$ required to ensure that there is no excess 
entropy generation at the edges $A_i$, thereby restricting the excess entropy production only at the GBs 
and the junction. The edges are allowed to carry excess entropy only when they are present
on the external surface of the solid, in which case the considered term will not be required. 
This additional power input will therefore be present only for edges in the interior 
of region $P$. Its form (and hence of ${\boldsymbol c}_i$) will of course depend on the 
constitutive nature of the prescribed excess quantities over GBs and the junction. It can be 
alternatively interpreted as the 
power expended by the configurational force ${\boldsymbol c}_i$ at the respective edge. 
This viewpoint has been adopted in earlier studies (cf. \cite{cermelli2,simha1,fried1}) within 
the framework of configuration mechanics. Our treatment (see also \cite{gupta1, basak1}) is 
different from these in that we do not introduce \textit{a priori} any balance law associated 
with configurational forces, nor do we postulate configurational forces as independent 
fundamental entities alongside the standard forces. It should be noted that the final
results are identical, irrespective of the chosen standpoint. We will now exploit the above 
restriction on the nature of entropy production, combined with certain constitutive restrictions on 
 GB energy, to determine ${\boldsymbol c}_i$. This will then be used to obtain local 
dissipation inequalities at various GBs and the junction.

Applying the transport theorem for an internal boundary (cf. Equation (A8) of \cite{simha1}) in \eqref{dissi_ineq_junc1} we obtain 
\begin{eqnarray}
\int_P\nabla\cdot({\boldsymbol\sigma}{\boldsymbol v})\,da+\sum_{i=1}^3\int_{\Gamma_i}\left([\![{\boldsymbol\sigma}
{\boldsymbol n}_i\cdot{\boldsymbol v}]\!]+f_i V_i-\mathring\gamma_i-
\frac{\partial}{\partial s_i}(\mu h_i)\right) dl+ \sum_{i=1}^3({\boldsymbol c}_i
\cdot{\boldsymbol w}_i-\gamma_i W_i)_{A_i}+\nonumber\\
\lim_{\epsilon\to 0}\oint_{{\mathscr C}_\epsilon}{\boldsymbol\sigma}
{\boldsymbol m}\cdot{\boldsymbol v}\,dl+\sum_{i=1}^3(\gamma_i{\boldsymbol t}_i
\cdot{\boldsymbol q}-\mu h_i)_J \geq 0,
\label{dissi_ineq_junc2}
\end{eqnarray} 
where $f_i=\gamma_i\kappa_i$ ($\kappa_i$ is the curvature of $\Gamma_i$), and $W_i$ is the 
tangential component of the edge velocity ${\boldsymbol w}_i$ at $A_i$. 
We consider isotropic GB energy such that $\gamma_i=\gamma_i(\theta_i)$, where 
$\theta_i$ is the misorientation angle at the boundary $\Gamma_i$. Using 
\eqref{bom_junc}$-$\eqref{blmjun_junc_stress} in \eqref{dissi_ineq_junc2}, then yields
\begin{eqnarray}
\int_P{\boldsymbol\sigma}\cdot\nabla{\boldsymbol v}\,da+\sum_{i=1}^3\int_{\Gamma_i}
\left([\![U_i{\boldsymbol E}]\!]{\boldsymbol n}_i \cdot{\boldsymbol n}_i+
\langle{\boldsymbol\sigma}{\boldsymbol n}_i\rangle\cdot[\![{\boldsymbol v}]\!]_t+f_i V_i-
\frac{\partial\gamma_i}{\partial\theta_i}\dot\theta_i-h_i\frac{\partial\mu}{\partial s_i}
\right) dl+\sum_{i=1}^3(\gamma_i{\boldsymbol t}_i\cdot{\boldsymbol q})_J+ \nonumber\\
\lim_{\epsilon\to 0}\oint_{{\mathscr C}_\epsilon}{\boldsymbol\sigma}{\boldsymbol m}\cdot{\boldsymbol v}\,dl
-\mu\lim_{\epsilon\to 0}\oint_{{\mathscr C}_\epsilon}\rho\,({\boldsymbol u}-{\boldsymbol v})\cdot{\boldsymbol m}
\,dl+\sum_{i=1}^3({\boldsymbol c}_i\cdot{\boldsymbol w}_i-\gamma_i W_i)_{A_i} \geq 0,
\label{dissi_ineq_junc23}
\end{eqnarray}
where $ {\boldsymbol E}=-(\rho\mu{\boldsymbol I}+{\boldsymbol\sigma})$ is the Eshelby tensor 
in the grains with vanishing bulk energy density (${\boldsymbol I}$ represents the identity tensor), $[\![{\boldsymbol v}]\!]_t$ is the tangential part 
of $[\![{\boldsymbol v}]\!]$, and we have used $\mathring\theta_i=\dot\theta_i$ recalling that 
the orientation of various grains remain uniform (since they are defect free).
 We have also imposed local chemical equilibrium at various boundaries, i.e. $[\![\mu]\!]=0$ 
 across $\Gamma_i$ \cite{fried1}.  
 The three summations in \eqref{dissi_ineq_junc23} represent the entropy production rate 
 associated with the GBs $\Gamma_i$,
 the junction $J$, and the edges $A_i$ lying on the boundary of the part, 
 respectively.
 We require the excess entropy production to have no contribution from the edges, hence 
 expecting it to be of the form $\sum_{i=1}^3\int_{\Gamma_i}\eta_i dl
 +\eta_J\geq 0$, where $\eta_i$ is the entropy generation rate per unit length of $\Gamma_i$
 and $\eta_J$ is the entropy generation rate at the junction. As a 
 consequence, we derive
\begin{equation}
{\boldsymbol c}_i=\gamma_i\,{\boldsymbol t}_i,
\label{cn_configforce}
\end{equation}
cf. Equations (17.4) and (17.21) in \cite{fried1}. The following local dissipation 
inequalities are then imminent
\begin{equation}
{\boldsymbol\sigma}\cdot\nabla{\boldsymbol v}\geq 0 \hspace{4mm}{\rm in}~P_i,
\label{local_inequality_grains_stress}
\end{equation}
\begin{equation}
[\![U_i{\boldsymbol E}]\!]{\boldsymbol n}_i\cdot{\boldsymbol n}_i+\langle{\boldsymbol\sigma}
{\boldsymbol n}_i\rangle\cdot[\![{\boldsymbol v}]\!]_t+f_i V_i-\frac{\partial\gamma_i}
{\partial\theta_i}\dot\theta_i-h_i\frac{\partial\mu}{\partial s_i}\geq 0 \hspace{4mm}
{\rm on}~\Gamma_i~\text{and}
\label{local_inequality_gb1}
\end{equation}
\begin{equation}
{\boldsymbol F}\cdot{\boldsymbol q}+\lim_{\epsilon\to 0}\oint_{{\mathscr C}_\epsilon}
{\boldsymbol E}{\boldsymbol m}\cdot({\boldsymbol q}-{\boldsymbol v})\,dl\geq 0 \,\,~{\rm at}~J,
\label{local_inequality_junc1}
\end{equation}
where 
\begin{equation}
{\boldsymbol F}=\sum_{i=1}^3 \gamma_i\,{\boldsymbol t}_i
\label{local_force_junc1}
\end{equation}
 is a part of the driving force for 
junction motion, cf. \cite{simha1}. The L.H.S. of these inequalities represent the dissipation 
rate in the grains (per unit area), at the boundaries (per unit length), and at the junction, 
respectively. Relation \eqref{local_inequality_grains_stress} requires the power expenditure
in the grains due to stress to be non-negative. When the grains are rigid, as is the 
case in this paper, the power expenditure in the bulk is identically zero and hence
\eqref{local_inequality_grains_stress} is trivially satisfied. Inequality 
\eqref{local_inequality_gb1} can be used to distinguish the fluxes (generalized velocities) and
the associated driving forces which cause
dissipation at a GB. Therefore the average traction drives the relative tangential jump in 
the velocity between two grains, the mean curvature drives
the normal velocity of the GB, and the torque like term  
$\partial\gamma_i/\partial\theta_i$ drives the evolution of the misorientation. The gradient
of the chemical potential acts as a driving 
force for mass diffusion along the GB. At the junction, according to 
\eqref{local_inequality_junc1}, we see that both GB energies of the intersecting boundaries and the singular Eshelby tensor 
in its neighbourhood contribute to the net dissipative force.

\section{Kinetics in a tricrystal}
\label{kineticsss_tricrystal_stress}
\begin{figure}[t!]
\centering
  \includegraphics[width=4.1in, height=2.4in] {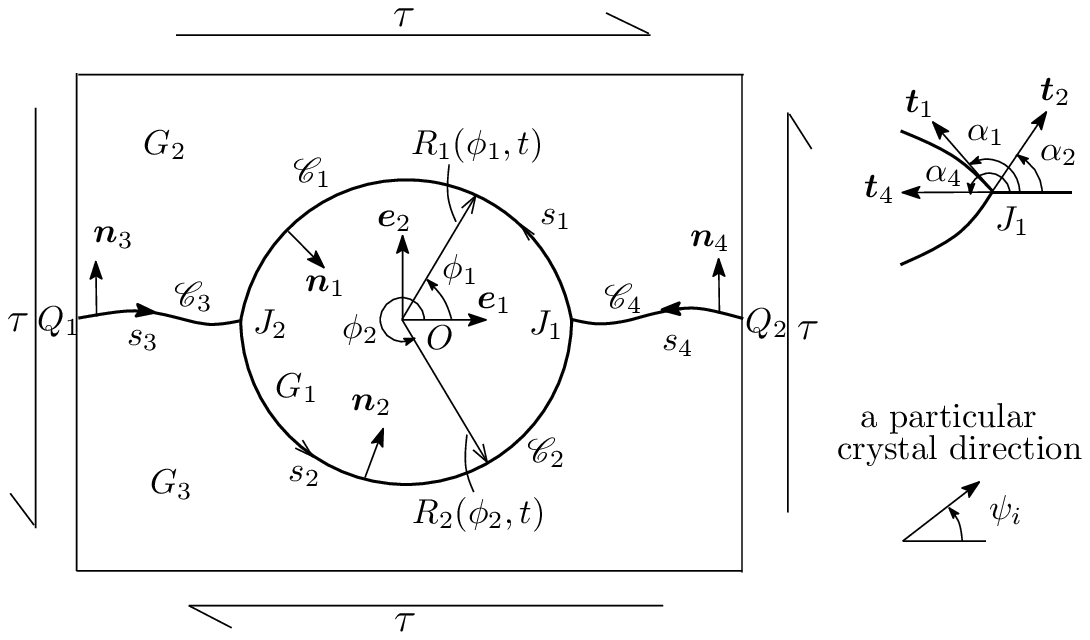}
\caption{\small{A schematic of the tricrystal. } }
\label{tricrystal_2D_stress}
\end{figure}
Based on the dissipation inequalities \eqref{local_inequality_gb1} and \eqref{local_inequality_junc1}
we now derive kinetic relations for 
a 2D tricrystal subjected to shear stress as shown in Figure \ref{tricrystal_2D_stress}. The tricrystal configuration 
consists of three grains $G_1$, $G_2$, and $G_3$, four GBs 
${\mathscr C}_i$, $i=1,\ldots,4$, and two junctions $J_1$ and $J_2$. The orientation of the respective 
grains, denoted by $\psi_1$, $\psi_2$ and $\psi_3$ (measured anticlockwise w.r.t. the ${\boldsymbol e}_1$-axis), 
are considered to be homogeneous (since they are all rigid and defect free). The tricrystal lies 
in a plane (spanned by 
${\boldsymbol e}_1$ and ${\boldsymbol e}_2$) orthogonal to ${\boldsymbol e}_3$, where 
$\{ {\boldsymbol e}_1, {\boldsymbol e}_2, {\boldsymbol e}_3 \}$ forms a right-handed orthonormal 
basis. The origin of the coordinate system is taken to coincide with 
the center of rotation of $G_1$. Since we assume that the tricrystal is initially 
symmetric about ${\boldsymbol e}_2$-axis, and also that the external loading is   
symmetric about the same axis, the instantaneous rotational velocity of two points in 
$G_1$ located in the neighborhood of $J_1$ 
and $J_2$ will always be equal and opposite until $G_1$ disappears. As a consequence, the 
mid-point of the line joining $J_1$ and $J_2$ will throughout represent the center of rotation,  
where the rotation axis is parallel to ${\boldsymbol e}_3$. Thus $\{ {\boldsymbol e}_1, {\boldsymbol e}_2,
{\boldsymbol e}_3 \}$ represents a basis for the translating coordinate with the origin held fixed with 
the instantaneous center of
rotation. The misorientation angles along ${\mathscr C}_1$, ${\mathscr C}_2$, and ${\mathscr C}_{3,4}$ 
are defined as $\theta_1=\psi_1-\psi_2$,
$\theta_2=\psi_1-\psi_3$, and $\theta_3=\psi_2-\psi_3$, respectively. The arc-length parameter 
for ${\mathscr C}_i$ is denoted by $s_i$ ($i=1,\ldots, 4)$ with an increasing  direction as shown in 
Figure \ref{tricrystal_2D_stress}. The normal ${\boldsymbol n}_i$ and the tangent ${\boldsymbol t}_i$ 
for a GB ${\mathscr C}_i$ is also shown in the same figure, where the latter is aligned in the direction 
of increasing $s_i$. 
The state of stress in each of the grains is considered to be given by 
\begin{equation}
{\boldsymbol\sigma}=\tau({\boldsymbol e}_1\otimes{\boldsymbol e}_2+{\boldsymbol e}_2\otimes
{\boldsymbol e}_1),
 \label{state_of_stress_bicrystal1}
\end{equation}
 which obviously satisfies the equilibrium equations \eqref{blm_grain_junc_stress} and 
 \eqref{blms_junc_stress} and the relevant traction boundary conditions. Let $R_i(\phi_i,t)$ be 
 the radial distance of the GB ${\mathscr C}_i$ from the center of rotation $O$ measured 
at an angle $\phi_i$ w.r.t. ${\boldsymbol e}_1$-axis (see Figure \ref{tricrystal_2D_stress}), where 
$0\leq \phi_1\leq\pi$ and $\pi\leq\phi_2\leq2\pi$. Denote the position vectors for the junctions $J_1$
and $J_2$ by ${\boldsymbol z}_1$ and ${\boldsymbol z}_2$, respectively and the position vectors
of the edges $Q_1$ and $Q_2$ by ${\boldsymbol z}_3$ and ${\boldsymbol z}_4$, respectively.

We allow 
the embedded grain $G_1$ to both rotate and translate with respect to the neighboring grains. 
The outer grains $G_2$ and 
$G_3$ are however restricted to undergo only relative translational motion.
 This is consistent with the observations made through MD simulations in \cite{trautt3}. 
 Without loss of generality we assume that $G_3$ remains stationary. 
Hence, $\dot\theta_1=\dot\theta_2=\dot\psi_1$ and $\dot\theta_3=0$. 
Based on these assumptions, the velocity of the three grains take the form
\begin{equation}
{\boldsymbol v}_1=\dot\psi_1{\boldsymbol e}_3\times{\boldsymbol x}+\dot{\boldsymbol C}_1,
\hspace{5mm}{\boldsymbol v}_2=\dot{\boldsymbol C}_2,\hspace{5mm}{\boldsymbol v}_3
={\boldsymbol 0},
\label{velocity_vector_grains_junc}
\end{equation}
where ${\boldsymbol C}_1$ and ${\boldsymbol C}_2$ are the rigid translations of $G_1$ and
$G_2$, respectively. Define vectors ${\boldsymbol{\mathcal C}}_1={\boldsymbol C}_1
-{\boldsymbol C}_2$, ${\boldsymbol{\mathcal C}}_2={\boldsymbol C}_1$
and ${\boldsymbol{\mathcal C}}_{3,4}={\boldsymbol C}_2$, as representing the relative 
translation between the adjacent grains across ${\mathscr C}_1$, ${\mathscr C}_2$, and 
${\mathscr C}_{3,4}$, respectively.
The normal and the tangent vector for ${\mathscr C}_a$ (from now on suffix $a$ will stand for
either $1$ or $2$, and suffix $b$ for either $3$ or $4$) can be written as
\begin{equation}
{\boldsymbol n}_a=-\frac{(R_a\cos\phi_a-{R'_a}\sin\phi_a)}{\sqrt{R_a^2+{R'_a}^2}}{\boldsymbol e}_1
-\frac{(R_a\sin\phi_a+{R'_a}\cos\phi_a)}{\sqrt{R_a^2+{R'_a}^2}}{\boldsymbol e}_2,~\text{and}
\label{normal_vector_gb1}
\end{equation}
\begin{equation}
{\boldsymbol t}_a=({\boldsymbol n}_a\cdot{\boldsymbol e}_2){\boldsymbol e}_1-({\boldsymbol n}_a
\cdot{\boldsymbol e}_1){\boldsymbol e}_2,
\label{tangent_vector_gb1}
\end{equation}
respectively (no summation for the repeated index $a$), where ${R'_a}=\partial R_a/\partial\phi_a$. 
Using ${\boldsymbol n}_a={\boldsymbol e}_3\times{\boldsymbol t}_a$
the normal component of the velocity ${\boldsymbol v}_1$, when evaluated on ${\mathscr C}_1$ and 
${\mathscr C}_2$, yields
\begin{equation}
{\boldsymbol v}_1\cdot{\boldsymbol n}_1=\dot\psi_1{\boldsymbol x}\cdot
{\boldsymbol t}_1+\dot{\boldsymbol C}_1\cdot{\boldsymbol n}_1~\text{and}~
{\boldsymbol v}_1\cdot{\boldsymbol n}_2=\dot\psi_1{\boldsymbol x}
\cdot{\boldsymbol t}_2+\dot{\boldsymbol C}_1\cdot{\boldsymbol n}_2,
\label{velocity_normal_grains_junc}
\end{equation}
respectively.

\paragraph{Consequence of mass balance}
Earlier MD  simulations \cite{trautt3} have confirmed that in the absence 
of external stress, the embedded grains spontaneously rotates about a fixed center of rotation 
and shrinks without any translational motion under GB 
capillary force. Based on this, and considering that the stress amplitude $\tau$ is much smaller
than the yield stress, we assume the translational velocities to be much smaller than the rotational 
velocity. As a result we ignore the effect of translational velocity on GB 
diffusion, and using \eqref{velocity_vector_grains_junc} and \eqref{velocity_normal_grains_junc}
in \eqref{boms_junc} we rewrite the mass balance at the GBs as 
\begin{eqnarray}
&& \frac{\partial h_a}{\partial s_a} = -\rho\,\dot{\psi_1}{\boldsymbol x}_a
\cdot{\boldsymbol t}_a 
~\text{on}~{\mathscr C}_a~\text{and}\nonumber\\
&& \frac{\partial h_b}{\partial s_b} = -\rho\,\dot{\boldsymbol{ C}}_2
\cdot{\boldsymbol n}_b~\text{on}~{\mathscr C}_b.
\label{balance_mass_2Djuncex}
\end{eqnarray}
Expanding \eqref{bomjun_junc} we obtain the following conditions for the diffusion currents at 
the junctions:
\begin{equation}
h_1-h_2-h_4=0 ~{\rm at}~ J_{1}~{\rm and}~h_1-h_2+h_3=0
~{\rm at}~J_{2},
\label{balance_mass_junctex}
\end{equation}
Substituting \eqref{normal_vector_gb1} and \eqref{tangent_vector_gb1} into  \eqref{balance_mass_2Djuncex}, while recalling that 
${\boldsymbol n}_b=(-1)^{b-1}{\boldsymbol e}_3\times{\boldsymbol t}_b$, and integrating the result yields
\begin{eqnarray}
&& h_a = \frac{\rho\dot\psi_1}{2}R_a^2 +k_a~\text{and} \nonumber\\
&& h_b=(-1)^{b}\rho\,\dot{\boldsymbol{C}}_2\cdot{\boldsymbol e}_3\times{\boldsymbol x}_b+k_b,
\label{mass_balance_tc_jun_stress2}
\end{eqnarray}
where $k_a$ and $k_b$ are the integration constants, and ${\boldsymbol x}_b={\boldsymbol x}(s_b)$.
Since the tricrystal has been assumed not to exchange mass with the surrounding, $h_b$ must satisfy
$h_3({\boldsymbol z}_3)=0$ and $h_4({\boldsymbol z}_4)=0$. With these boundary
conditions, \eqref{mass_balance_tc_jun_stress2}$_2$ simplifies to
\begin{equation}
h_b=(-1)^b\rho\,\dot{\boldsymbol{C}}_2\cdot{\boldsymbol e}_3\times({\boldsymbol x}_b-{\boldsymbol z}_b).
\label{mass_balance_tc_jun_stress3}
\end{equation}
The GBs ${\mathscr C}_3$ and ${\mathscr C}_4$ are always symmetrically equivalent about 
${\boldsymbol e}_2$-axis, hence the validity of \eqref{mass_balance_tc_jun_stress3} for both these
curves demands that $\dot{\boldsymbol{C}}_2$ must be 
parallel to ${\boldsymbol e}_1$, i.e. $\dot{\boldsymbol{C}}_2=\dot{C}_2{\boldsymbol e}_1$, which
implies that the upper grain will always move in a horizontal direction with respect to the lower
grains. Consequently,
\eqref{mass_balance_tc_jun_stress3} reduces down to
\begin{equation}
h_b=(-1)^{b-1}\rho\,\dot{C}_2\,y_b,
\label{mass_balance_tc_jun_stress4}
\end{equation}
where $y_b=({\boldsymbol x}_b-{\boldsymbol z}_b)\cdot{\boldsymbol e}_2$.
Using \eqref{mass_balance_tc_jun_stress2}$_1$ and
\eqref{mass_balance_tc_jun_stress4} in \eqref{balance_mass_junctex} and neglecting the contribution
coming from the translational 
velocity in comparison to the rotational speed as far as the diffusion fluxes along 
${\mathscr C}_1$ and 
${\mathscr C}_2$ are concerned, we conclude that 
\begin{equation}
k_1-k_2=0 ~ \text{at}~J_1~\text{and}~J_2.
\label{mass_balance_tc_jun_stress32}
\end{equation}
 Equations \eqref{mass_balance_tc_jun_stress32} and \eqref{mass_balance_tc_jun_stress2}$_1$
in association with the conservation condition $\int_{{\mathscr C}_1} h_1dl
+\int_{{\mathscr C}_2}h_2dl=0$ along the closed boundary of $G_1$ yield 
\begin{equation}
h_a=\frac{\rho\dot\psi_1}{2}(R^2-\overline{R^2}),
\label{mass_balance_tc_jun_stress5}
\end{equation}
where $\overline{R^2}=(\int_{{\mathscr C}_1}R_1^2\,dl+\int_{{\mathscr C}_2}R_2^2\,dl)
/(|{\mathscr C}_1|+|{\mathscr C}_2|)$ and $|{\mathscr C}_i|$ is the length of ${\mathscr C}_i$. 
The conservation condition can be readily proved using the Fick's law
\begin{equation}
h_i=-D_i\frac{\partial\mu}{\partial s_i},
\label{ficks_law_tricrystal_stress}
\end{equation}
where $D_i\geq 0$ is the diffusivity along the GB ${\mathscr C}_i$. Moreover, applying 
\eqref{ficks_law_tricrystal_stress} in \eqref{mass_balance_tc_jun_stress4} and then integrating 
the equation we obtain the chemical potential along ${\mathscr C}_b$ as
\begin{equation}
\mu_b=(-1)^b\frac{\rho\,\dot{C}_2}{D_b}I_b, 
\label{chemical_potential_gb34}
\end{equation}
where $I_b=\displaystyle\int_{s_b({\boldsymbol z}_b)}^{s_b}y_b\,dl$. The chemical potential at 
the free surfaces (assumed to be flat) has been considered to be zero as there are no normal 
components of traction on those faces (cf. \cite{herring2} and Chapter 68 in  \cite{gurtin_cont_mech}).

\paragraph{Grain and GB kinetics}
We begin by deriving the kinetic laws relevant to ${\mathscr C}_a$ and $G_1$ before moving on to the kinetics of other GBs and junctions.
Ignoring the terms of the order of $\dot{\mathcal C}_a\dot\psi_1$ and $\dot{\mathcal C}_a^2$, 
while using \eqref{blms_junc_stress}, 
\eqref{velocity_vector_grains_junc}, \eqref{velocity_normal_grains_junc}, and 
\eqref{ficks_law_tricrystal_stress} in \eqref{local_inequality_gb1}, 
we rewrite the dissipation inequality on ${\mathscr C}_a$ as
\begin{equation}
f_aV_a+g_a\nu_a+\sigma_a\dot{\mathcal C}_a\geq 0,
\label{dissi_ineq_tricrystal2}
\end{equation}
where $\dot{\mathcal C}_a=(\dot{\boldsymbol C}_1-\dot{\boldsymbol C}_2)\cdot{\boldsymbol e}_1$, 
$\bar{D}_a=D_a/\rho^2$, and
 \begin{equation}
g_a=\frac{1}{{\boldsymbol x}_a\cdot{\boldsymbol n}_a}\left(\frac{\partial\gamma_a}
{\partial\theta_a}-\rho\mu_a{\boldsymbol x}_a\cdot{\boldsymbol t}_a
-\frac{{\dot\psi_1}}{4\bar{D}_a}(R_a^2-\overline{R_a^2})^2-T_a\right) 
\label{forces_tricrystal2}
\end{equation}
is the driving force for the rotational motion of $G_1$ with
\begin{equation}
 T_a=\tau(({\boldsymbol n}_a\cdot{\boldsymbol e}_1)({\boldsymbol x}_a\cdot{\boldsymbol e}_1)-
 ({\boldsymbol n}_a\cdot{\boldsymbol e}_2)({\boldsymbol x}_a\cdot{\boldsymbol e}_2)); \nonumber
\label{forces_tricrystal222}
\end{equation}
$\nu_a=-\dot\psi_1\,{\boldsymbol x}_a\cdot{\boldsymbol n}_a$ is related to the rotational velocity of 
$G_1$; and $\sigma_a=\langle{\boldsymbol\sigma}{\boldsymbol n}_a\rangle\cdot{\boldsymbol e}_1=\tau
{\boldsymbol n}_a\cdot{\boldsymbol e}_2$ is the driving force for the translational motion between
the adjacent grains.
While deriving \eqref{dissi_ineq_tricrystal2} we have neglected a term proportional to 
$\mu_a\dot{\mathcal C}_a$, which is estimated to be of the order of $\dot\psi_1\dot{\mathcal C}_a 
+\dot{\mathcal C}_a^2$ considering \eqref{ficks_law_tricrystal_stress} and 
\eqref{mass_balance_tc_jun_stress5}.
The term $\sigma_a\dot{\mathcal C}_a$ in \eqref{forces_tricrystal2} is derived from what originally 
was of the form $\langle{\boldsymbol\sigma}{\boldsymbol n}_a
\rangle\cdot\dot{\boldsymbol{\mathcal C}}_a$. Indeed, the tricrystal will always maintain the symmetry (about 
${\boldsymbol e}_2$-axis) as dictated by its initial geometry and the loading condition.
Moreover, since the relative velocities $\dot{\boldsymbol{\mathcal C}}_a$ are uniform
over the respective GBs, we can justifiably assume that it is only the average values of their conjugate
forces, i.e. $\int_{{\mathscr C}_a}\tau (n_2{\boldsymbol e}_1+n_1{\boldsymbol e}_2)dl/
|{\mathscr C}_a|$, which
is ultimately going to contribute to 
the net dissipation.

Assuming linear kinetics, and recalling the Onsager's reciprocity theorem, 
we consider the following set of 
phenomenological kinetic equations for the fluxes on ${\mathscr C}_a$ \cite{cahn1,basak1}:
\begin{equation}
V_a = M_af_a+ M_a\beta_ag_a, 
\label{kinetics_tricrystal5}
\end{equation}
\begin{equation}
\nu_a = \beta_aV_a+S_ag_a,~\text{and}
\label{kinetics_tricrystal4}
\end{equation}
\begin{equation}
\dot{{\mathcal C}}_a = L_a\sigma_a,
\label{kinetics_tricrystal3}
\end{equation}
where $M_a>0$, $\beta_a$, $S_a\geq 0$, and $L_a\geq 0$ are the mobility, geometric coupling factor,
viscous sliding coefficient, and translational coefficient associated with ${\mathscr C}_a$. 
The restrictions on $M_a$, $S_a$, and $L_a$ can be easily verified by using 
\eqref{kinetics_tricrystal5}, \eqref{kinetics_tricrystal4}, and \eqref{kinetics_tricrystal3}
in the inequality \eqref{dissi_ineq_tricrystal2}. For the same reason as discussed above (see the 
paragraph following \eqref{velocity_normal_grains_junc}), we have
assumed that the translational velocity of the embedded grain is decoupled from rotational evolution 
and migration. Multiplying both sides of \eqref{kinetics_tricrystal5} by $S_a$ and then replacing 
$g_a$ from it using \eqref{kinetics_tricrystal4} we get
\begin{equation}
V_a=\frac{M_a}{S_a+M_a\beta_a^2}(S_af_a+\beta_a\nu_a),
\label{GB_velocity_tricrystala}
\end{equation}
which is the governing equation for the normal velocity of ${\mathscr C}_a$. 
To calculate $\dot\psi_1$ we begin by combining \eqref{kinetics_tricrystal5} with 
\eqref{kinetics_tricrystal4}, after replacing $g_a$ from \eqref{forces_tricrystal2}, to obtain
\begin{equation}
\dot\psi_1\left(\frac{({\boldsymbol x}_a\cdot{\boldsymbol n}_a)^2}{S_a+M_a\beta_a^2}
-\frac{1}{4\bar{D}_a}(R_a^2-\overline{R_a^2})^2\right)=-\frac{M_a\beta_a}{S_a+M_a\beta_a^2}
({\boldsymbol x}_a\cdot{\boldsymbol n}_a)f_a-\frac{\partial\gamma_a}
{\partial\theta_a}+T_a+\mu_a{\boldsymbol x}_a\cdot{\boldsymbol t}_a.
\label{grain_rotation_tricrystala1}
\end{equation}
These two equations (for $a=1,2$) are then integrated over ${\mathscr C}_1$ and 
${\mathscr C}_2$, respectively, and summed up to write the ordinary differential equation 
\begin{equation}
\dot\psi_1=\frac{-\displaystyle\sum_{a=1}^2\int_{{\mathscr C}_a}
\left(\displaystyle\frac{M_a\beta_a}{S_a+M_a\beta_a^2}f_a\,{\boldsymbol x}_a\cdot
{\boldsymbol n}_a+\frac{\partial\gamma_a}{\partial\theta_a}-T_a\right)dl}{
\displaystyle\sum_{a=1}^2\int_{{\mathscr C}_a}\left(\displaystyle\frac{({\boldsymbol x}_a\cdot
{\boldsymbol n}_a)^2}{S_a+M_a\beta_a^2}-\frac{1}{2\bar{D}_a}(R_a^2-\overline{R_a^2})^2\right)dl},
\label{grain_rotation_tricrystala2}
\end{equation} 
where we have used the identity
(obtained using ${\boldsymbol t}_a=d{\boldsymbol x}_a/ds_a$, \eqref{ficks_law_tricrystal_stress} 
and \eqref{mass_balance_tc_jun_stress5})
\begin{equation}
\sum_{a=1}^2\int_{{\mathscr C}_a}\mu_a{\boldsymbol x}_a
\cdot{\boldsymbol t}_adl=\sum_{a=1}^2\frac{\dot\psi_1}{4\bar{D}_a}(R_a^2-\overline{R_a^2})^2dl
\nonumber.
\label{grain_rotation_tricrystala41}
\end{equation}
We now have the kinetic relations governing the normal velocities of ${\mathscr C}_1$ 
and ${\mathscr C}_2$ in \eqref{GB_velocity_tricrystala}, and the rotational speed of the inner 
grain $G_1$ in \eqref{grain_rotation_tricrystala2}. In the following we will derive kinetics for 
the normal velocity of boundaries ${\mathscr C}_3$ and ${\mathscr C}_4$,  the translational velocity of the 
grains $G_1$ and  $G_2$, and  the velocities for $J_1$ and $J_2$. 

The dissipation inequality \eqref{local_inequality_gb1} for GBs 
${\mathscr C}_3$ and ${\mathscr C}_4$, across which there is no misorientation evolution, can 
be reduced to 
\begin{equation}
f_b\,V_b+\sigma_b\,(\dot{C}_2)_b\geq 0,
\label{dissi_ineq_tricrystal_gb34}
\end{equation}
where 
\begin{equation}
 \sigma_b=(-1)^b\frac{(\dot{C}_2)_b}{\bar{D}_b}I_b{\boldsymbol n}_b\cdot{\boldsymbol e}_1
+\tau{\boldsymbol n}_b\cdot{\boldsymbol e}_2+\frac{\rho^2}{D_b}
(\dot{C}_2)_b y_b^2.
\label{driving_forces_tricrystala_gb34}
\end{equation}
As done previously we postulate linear kinetics 
from \eqref{dissi_ineq_tricrystal_gb34}:
\begin{equation}
(\dot{C}_2)_b=\beta_b V_b+L_b\,\sigma_b~\text{and}
\label{kinetic_tricrystal_planar1}
\end{equation}
\begin{equation}
V_b=M_b\,f_b+M_b\,\beta_b\, \sigma_b.
\label{kinetic_tricrystal_planar2}
\end{equation}
To derive an expression for the average translational velocity of $G_2$ we substitute $V_b$ from 
\eqref{kinetic_tricrystal_planar2} 
into \eqref{kinetic_tricrystal_planar1}, then integrate 
it over ${\mathscr C}_3$ and ${\mathscr C}_4$, respectively for $b=3,4$, and finally add 
them up to obtain
\begin{equation}
\dot{C}_2
=\frac{\displaystyle\sum_{b=3}^4\displaystyle\int_{\mathscr{C}_b}\left(\frac{M_b\beta_b}
{L_b+M_b\beta_b^2}\,f_b
+\tau{\boldsymbol n}_b\cdot{\boldsymbol e}_2\right)dl}
{\displaystyle\sum_{b=3}^4\displaystyle\int_{\mathscr{C}_b}\left(\frac{1}{L_b+M_b\beta_b^2}
-(-1)^{b}\frac{I_b}{\bar{D}_b}{\boldsymbol n}_b\cdot{\boldsymbol e}_1-\frac{1}{\bar{D}_b}
y_b^2\right)dl},
\label{kinetic_tricrystal_planar4} 
\end{equation}
where $M_b$, $\beta_b$, $S_b$, and $L_b$ have the same meaning as described above.
Eliminating $\sigma_b$ between \eqref{kinetic_tricrystal_planar2} and 
\eqref{kinetic_tricrystal_planar1} we obtain the governing kinetic law for the normal velocity of
${\mathscr C}_b$ as
\begin{equation}
V_b=\frac{M_bL_b}{L_b+M_b\beta_b^2}f_b+\frac{M_b\beta_b}{L_b+M_b\beta_b^2}\dot{C}_2,
\label{kinetic_tricrystal_planar311} 
\end{equation}
where we have replaced $(\dot{C}_2)_b$ by the average translational rate of $G_2$ given by
\eqref{kinetic_tricrystal_planar4}.
Next we integrate \eqref{kinetics_tricrystal3} for $a=1$ and $2$, respectively,
and combine them to obtain the average of the translational velocity of $G_1$
\begin{equation}
\dot{C}_1= \frac{1}{|{\mathscr C}_1|+|{\mathscr C}_2|}\left({\dot{C}_2|{\mathscr C}_1|}+ \sum_{a=1}^{2}
\int_{{\mathscr C}_a}L_a\sigma_adl\right).
\label{grain_rotation_tricrystala42}
\end{equation}  
We now have all the required kinetic relations related to GB motion and grain dynamics. 

\paragraph{Junction kinetics}
We will next derive the kinetic relations for the 
two junctions. Using \eqref{velocity_vector_grains_junc}
and the weak singularity in the stress field (see \eqref{blmjun_junc_stress} and the discussion in 
Appendix \ref{balance_laws_stress_2D1}) , in addition to assuming $\rho$ to be nonsingular, one can easily show that
the closed integral in \eqref{local_inequality_junc1} would vanish in the limit $\epsilon\to 0$, simplifying it to
\begin{equation}
{\boldsymbol F}_\delta\cdot{\boldsymbol q}_\delta\geq 0 ~{\rm at}~J_\delta, ~\text{for}~\delta=1,2.
\label{local_inequality_junc111}
\end{equation}
Linear kinetic relations can then be motivated from \eqref{local_inequality_junc111} as
 \cite{fischer1} 
\begin{equation}
{\boldsymbol q}_\delta=m_\delta{\boldsymbol F}_\delta ~\text{at}~J_\delta,
\label{junction_tricrystal_dynamics1}
\end{equation}
where $m_\delta\geq 0$ is the mobility coefficient associated with junction $J_\delta$,
\begin{equation}
{\boldsymbol F}_1=\gamma_1{\boldsymbol t}_1-\gamma_2{\boldsymbol t}_2
-\gamma_4{\boldsymbol t}_4,~\text{and}
\label{junction_force1_tricrystal}
\end{equation}
\begin{equation}
{\boldsymbol F}_2=-\gamma_1{\boldsymbol t}_1+\gamma_2{\boldsymbol t}_2
-\gamma_3{\boldsymbol t}_3.
\label{junction_force2_tricrystal}
\end{equation}
We assume the junctions to be non-splitting. Compatibility at the junctions would then require 
\cite{fischer1}
\begin{equation}
V_i={\boldsymbol q}_1\cdot{\boldsymbol n}_i ~\text{at}~J_1~\text{for}
~i=1,2,4,
\label{junction_tricrystal_compatibility1}
\end{equation}
\begin{equation}
V_i={\boldsymbol q}_2\cdot{\boldsymbol n}_i ~\text{at}~J_2~\text{for}
~i=1,2,3.
\label{junction_tricrystal_compatibility2}
\end{equation}
These compatibility equations will be used to determine the junction angles, as described below.
 
It follows from the geometry of the tricrystal that
\begin{equation}
{\boldsymbol n}_i\cdot{\boldsymbol t}_j=\sin(\alpha_j-\alpha_i) ~\text{and}~{\boldsymbol t}_i
\cdot{\boldsymbol t}_j=\cos(\alpha_i-\alpha_j)~\text{for}~i,j=1,2,3,4
\label{junction_angles_stress}
\end{equation}
 at the junctions, where $\alpha_i$ is the angle made by the tangent (in the limiting sense) to 
 ${\mathscr C}_i$  with ${\boldsymbol e}_1$-axis at the corresponding junction (see Figure 
 \ref{tricrystal_2D_stress}).
Substituting ${\boldsymbol F}_\delta$ from \eqref{junction_force1_tricrystal} and
\eqref{junction_force2_tricrystal}, using \eqref{junction_tricrystal_dynamics1}, 
in \eqref{junction_tricrystal_compatibility1} and \eqref{junction_tricrystal_compatibility2},
and then employing \eqref{junction_angles_stress} in the resulting relations we obtain the 
following sets of compatibility relations: 
\begin{eqnarray}
&& -\gamma_2\sin(\alpha_2-\alpha_1)-\gamma_4\sin(\alpha_4-\alpha_1)=\frac{V_1}{m_1}, \nonumber\\
&& \gamma_1\sin(\alpha_1-\alpha_2)-\gamma_4\sin(\alpha_4-\alpha_2)=\frac{V_2}{m_1}, \nonumber\\
&& \gamma_1\sin(\alpha_1-\alpha_4)-\gamma_2\sin(\alpha_2-\alpha_4)=\frac{V_4}{m_1} ~
\text{at}~J_1, ~\text{and}
\label{junction_compatibility_j1}
\end{eqnarray}
\begin{eqnarray}
&&\gamma_2\sin(\alpha_2-\alpha_1)-\gamma_3\sin(\alpha_3-\alpha_1)=\frac{V_1}{m_2}, \nonumber\\
&&-\gamma_1\sin(\alpha_1-\alpha_2)-\gamma_3\sin(\alpha_3-\alpha_2)=\frac{V_2}{m_2}, \nonumber\\
&&-\gamma_1\sin(\alpha_1-\alpha_3)+\gamma_2\sin(\alpha_2-\alpha_3)=\frac{V_3}{m_2}~
\text{at}~J_2,
\label{junction_compatibility_j2}  
\end{eqnarray}
when $m_\delta>0$ in finite. The nonlinear algebraic equations given by 
\eqref{junction_compatibility_j1} and \eqref{junction_compatibility_j2} have to be solved in 
order to obtain $\{\alpha_1,\alpha_2,\alpha_4\}$ and 
$\{\alpha_1,\alpha_2,\alpha_3\}$ at $J_1$ and $J_2$, respectively.
Using \eqref{junction_force1_tricrystal} and 
\eqref{junction_force2_tricrystal} in \eqref{junction_tricrystal_dynamics1} the junction
velocities are calculated as
\begin{eqnarray}
&&{\boldsymbol q}_1 = m_1(\gamma_1\cos\alpha_1-\gamma_2\cos\alpha_2-\gamma_4\cos\alpha_4)
{\boldsymbol e}_1+m_1(\gamma_1\sin\alpha_1-\gamma_2\sin\alpha_2-\gamma_4\sin\alpha_4)
{\boldsymbol e}_2 ~\text{and} \nonumber\\
&&{\boldsymbol q}_2 = m_2(-\gamma_1\cos\alpha_1+\gamma_2\cos\alpha_2-\gamma_3\cos\alpha_3)
{\boldsymbol e}_1+m_2(-\gamma_1\sin\alpha_1+\gamma_2\sin\alpha_2-\gamma_3\sin\alpha_3)
{\boldsymbol e}_2.
 \label{junction_velocity12_stress}
\end{eqnarray}
When the junction mobility is infinite, i.e. $m_\delta\to\infty$, 
\eqref{junction_compatibility_j1} and \eqref{junction_compatibility_j2} give two independent 
equations 
\begin{equation}
\frac{\gamma_1}{\sin(\alpha_2-\alpha_4)}=\frac{-\gamma_2}{\sin(\alpha_4-\alpha_1)}
=\frac{-\gamma_4}{\sin(\alpha_1-\alpha_2)} \hspace{3mm}\text{at}~J_1~\text{and}
\label{junction_compatibility_j11}
\end{equation}
\begin{equation}  
\frac{\gamma_1}{\sin(\alpha_2-\alpha_3)}=\frac{-\gamma_2}{\sin(\alpha_3-\alpha_1)}
=\frac{\gamma_3}{\sin(\alpha_1-\alpha_2)}~\text{at}~J_2,
\label{junction_compatibility_j12}
\end{equation}
known as the Young-Dupr\'{e} equations \cite{fischer1}. In order to solve the 
junction angles uniquely, we use the following equations which are obtained by eliminating 
$m_1$ and $m_2$ from the respective sets of equations from \eqref{junction_compatibility_j1} 
and \eqref{junction_compatibility_j2}:
\begin{equation}
\gamma_1V_1-\gamma_2V_2-\gamma_4V_4=0~\text{at}~J_1~\text{and}~
\gamma_1V_1-\gamma_2V_2+\gamma_3V_3=0~\text{at}~J_2.
\label{junction_compatibility_j13}
\end{equation}
To calculate the velocity of $J_1$ when $m_1\to\infty$, we write ${\boldsymbol q}_1=q_1
(\cos\xi{\boldsymbol e}_1+\sin\xi{\boldsymbol e}_2)$ where $\xi$ is the angle made by 
${\boldsymbol q}_1$ with ${\boldsymbol e}_1$-axis. Using this expression in 
\eqref{junction_tricrystal_compatibility1} twice (i.e. for two different values of $i$)
we get
\begin{equation}
{\boldsymbol q}_1=\csc(\alpha_j-\alpha_i)\left((V_i\cos\alpha_j-V_j\cos\alpha_i)
{\boldsymbol e}_1+(V_i\sin\alpha_j-V_j\sin\alpha_i){\boldsymbol e}_2\right)~
\text{for any}~i,j=1,2,4,\,i\neq j.
\label{junction_velocity_stress_minfty}
\end{equation}
The expression for the velocity of $J_2$ is same as \eqref{junction_velocity_stress_minfty},
except that the indices are now restricted to $i,j=1,2,3$.  

To summarize, the migration kinetics of GBs ${\mathscr C}_1$ and ${\mathscr C}_2$ is governed by 
\eqref{GB_velocity_tricrystala}, while those of ${\mathscr C}_3$ and 
${\mathscr C}_4$ by \eqref{kinetic_tricrystal_planar311}; all of these are non-linear parabolic partial differential equations.
Relations \eqref{grain_rotation_tricrystala2}, \eqref{kinetic_tricrystal_planar4}, and 
\eqref{grain_rotation_tricrystala42} govern 
the homogeneously evolving orientation for $G_1$, the uniform horizontal translation for $G_2$, and the
uniform horizontal translation for $G_1$, respectively, while keeping
grain $G_3$ fixed. The junction dynamics at $J_1$ and $J_2$ follow \eqref{junction_velocity12_stress}
when the junction mobility is finite. The unknown junction angles $\alpha_1$, $\alpha_2$,
and $\alpha_3$ at $J_1$ and $J_2$ 
are then obtained by solving the set of equations \eqref{junction_compatibility_j1} and 
\eqref{junction_compatibility_j2}.
The junction motion is governed by \eqref{junction_velocity_stress_minfty} 
when the mobility coefficient takes an infinite value; the three unknown junction angles are then 
determined by solving \eqref{junction_compatibility_j11}, \eqref{junction_compatibility_j12}, and 
\eqref{junction_compatibility_j13}.

\paragraph{Remark 1.} We consider a special case of the arrangement shown in 
Figure \ref{tricrystal_2D_stress} where the embedded grain is absent. We get a bicrystal with
two rectangular grains which are separated by a non-planar smooth GB (denote it by ${\mathscr C}$). Without loss of generality we assume the lower grain to be stationary, so that 
the kinetic equations for GB migration and sliding rate $\dot{C}$ of the upper grain can be obtained from 
\eqref{kinetic_tricrystal_planar311} and \eqref{kinetic_tricrystal_planar4} as
\begin{equation}
V=\frac{M L}{L+M\beta^2}f+\frac{M\beta}{L+M\beta^2}\dot{C} ~\text{and}
\label{kinetic_tricrystal_planar312} 
\end{equation}
\begin{equation}
\dot{C}=\frac{\displaystyle\displaystyle\int_{\mathscr{C}}\left(\frac{M\beta}{L+M\beta^2}\,f
+\tau{\boldsymbol n}\cdot{\boldsymbol e}_2\right)dl}
{\displaystyle\displaystyle\int_{\mathscr{C}}\left(\frac{1}{L+M\beta^2}
+\frac{I}{\bar{D}}{\boldsymbol n}\cdot{\boldsymbol e}_1-\frac{y^2}{\bar{D}}\right)dl},
\label{kinetic_tricrystal_planar411} 
\end{equation}
respectively, where all the symbols have the same meaning as before. 

When GB ${\mathscr C}$ is planar, $f=0$, ${\boldsymbol n}={\boldsymbol e}_2$, $y=0$, and $I=0$. As a result,
\eqref{kinetic_tricrystal_planar312} and \eqref{kinetic_tricrystal_planar411} simplify to
\begin{equation}
 V=M\beta\tau ~\text{and}~\dot{C}=(L+M\beta^2)\tau, 
\label{kinetic_tricrystal_planar_sols}
\end{equation}
respectively. These
are in agreement with the earlier work on low angle tilt GBs \cite{read2, molodov1} (where it is additionally assumed that $L=0$). On the other hand, when the GB is curved, but assumed to migrate without coupling and sliding, then the well known kinetic relations, $V=Mf$ and $\dot{C}=0$, are readily obtained.

As another scenario, consider GB migration to be absent so that only GB diffusion accommodated tangential motion of the grain is present. The non-steady-state sliding velocity, obtained from\eqref{kinetic_tricrystal_planar411}, is then governed by 
\begin{equation}
\dot{C}=\frac{\displaystyle\displaystyle\int_{\mathscr{C}}\tau{\boldsymbol n}\cdot{\boldsymbol e}_2\,dl}
{\displaystyle\displaystyle\int_{\mathscr{C}}\left(\frac{1}{L}+\frac{I}{\bar{D}}\,{\boldsymbol n}\cdot{\boldsymbol e}_1
-\frac{y^2}{\bar{D}}\right)dl}.
\label{kinetic_tricrystal_planar412} 
\end{equation}
Similar relations are used to model viscous GB sliding to understand creep \cite{raj1}.

\paragraph{Remark 2.} \label{remark2}
As another special case, we consider the arrangement shown in Figure \ref{tricrystal_2D_stress} without the non-planar 
GBs ${\mathscr C}_3$ and ${\mathscr C}_4$. We then have a bicrystal where a non-circular 
cylindrical grain is embedded inside another grain. Let us denote the closed  
GB curve by ${\mathscr C}$ and assume that it is 
smooth. Considering the outer grain of the bicrystal to be stationary, the kinetic 
equation for GB motion, grain rotation, and the translational rate of the embedded grain can be 
obtained from \eqref{GB_velocity_tricrystala}, \eqref{grain_rotation_tricrystala2}, and 
\eqref{grain_rotation_tricrystala42} as
\begin{equation}
V=\frac{M}{S+M\beta^2}(Sf+\beta\nu),
\label{GB_velocity_tricrystalaaa}
\end{equation}
\begin{equation}
\dot\theta=\frac{-\displaystyle\int_{{\mathscr C}}
\left(\displaystyle\frac{M\beta}{S+M\beta^2}f\,{\boldsymbol x}\cdot
{\boldsymbol n}+\frac{\partial\gamma}{\partial\theta}-T\right)dl}{
\displaystyle\int_{{\mathscr C}}\left(\displaystyle\frac{({\boldsymbol x}\cdot
{\boldsymbol n})^2}{S+M\beta^2}-\frac{1}{2\bar{D}}(R^2-\overline{R^2})^2\right)dl},~\text{and}
\label{grain_rotation_tricrystala2aa}
\end{equation} 
\begin{equation}
\dot{C}= \frac{1}{|{\mathscr C}|}\int_{{\mathscr C}}L\sigma\, dl,
\label{grain_rotation_tricrystala42aa}
\end{equation} 
respectively. In the absence of translational velocity 
of the embedded grain, i.e. $\dot{C} = 0$, the system of equations 
\eqref{GB_velocity_tricrystalaaa}-\eqref{grain_rotation_tricrystala2aa} coincide with the results 
derived in \cite{basak1,taylor1}. The above equations provide an extension to the previous work so as to not restrict the center of rotation of the embedded grain to be fixed.

\section{Results and discussion}
\label{results_junction_2D}
We introduce non-dimensional position and time variables as $\tilde{\boldsymbol x}={\boldsymbol x}/R_0$
and $\tilde{t}=t/t_0$, respectively, where we choose $t_0$ to be is the time taken for an isolated 
circular GB of radius $R_0$, with energy $\gamma_0$ and mobility $M_0$, to vanish under curvature 
driven migration, and hence $t_0=R_0^2/2\gamma_0M_0$. These dimensionless variables can be substituted 
in \eqref{GB_velocity_tricrystala}, \eqref{kinetic_tricrystal_planar311}, 
\eqref{grain_rotation_tricrystala2}, \eqref{grain_rotation_tricrystala42}, 
\eqref{kinetic_tricrystal_planar4}, and 
\eqref{junction_compatibility_j1}$-$\eqref{junction_velocity_stress_minfty}, 
to obtain a system of non-dimensionalized kinetic equations for the tricrystal. 
 This naturally introduces three dimensionless parameters $r_1=S_0/M_0$, $r_2=M_0R_0^2/\bar{D}$, 
 and $r_3=L_0/M_0$  associated with GB kinetics, 
and one non-dimensional parameter $\Lambda_\delta=2R_0m_\delta/M_0$  with junction 
kinetics \cite{basak1,czubayko1}. We restrict our simulations to constant mobility, sliding 
coefficients, and translation coefficient assumed to be same for all the GBs, and also constant junction mobility 
coefficient, considered same for both the junctions. Hence, say $\Lambda_1=\Lambda_2 =\Lambda$.
On the other hand, we consider an isotropic GB energy and a coupling factor as described by the solid 
curves in Figures 2(a) and 4(b), respectively, in \cite{basak1}. The values of the dimensionless parameters 
are taken as $0.01\leq r_1\leq 1$, $r_2=10^3({\overline{R}}(0)/{\overline{R}}
(\tilde{t}))^{3/2}$, and $1\leq\Lambda\leq \infty$ \cite{basak1,czubayko1}. 
The time-dependent term in $r_2$ ensures that with decreasing grain size GB diffusivity increases 
\cite{chen1}. Because of lack of proper data related to the translational coefficient $L_0$, we 
consider $r_3=1$ (unless stated otherwise) in order to observe tangible 
grain translations. All the parameters have been taken for face-centered cubic crystals.

The non-dimensionalized kinetic equations 
are solved numerically to investigate the shape and orientation evolution of the embedded
grain. Our simulation methodology is based on the finite difference scheme proposed by Fischer
et al. \cite{fischer1}. The scheme is now described briefly for the tricrystal.  
 The GB ${\mathscr C}_i$ (recall that the subscript $i$ refers to one of the GBs in the 
tricrystal arrangement) has been discretized with $N_i$ number of grid points
(discretization goes in the direction of increasing $s_i$)
with the position vector (non-dimensionalized) denoted by $\tilde{\boldsymbol x}_i^{j}=x_i^{j}
{\boldsymbol e}_1+y_i^{j}{\boldsymbol e}_2$, where the superscript $j=0,\ldots,N_i$ denotes the 
index of
the grid point on ${\mathscr C}_i$. The position of the grid points of ${\mathscr C}_a$ 
and ${\mathscr C}_b$ are updated at time 
instance $\tilde{t}_{n+1}$ using the following explicit time integration scheme (superposed tilde 
represents dimensionless variables):
\begin{equation}
 \tilde{\boldsymbol x}_a^{j}(\tilde{t}_{n+1})=\tilde{\boldsymbol x}_a^{j}(\tilde{t}_n)
 +\Delta\tilde{t}\,\tilde{V}_a^j(\tilde{t}_n)\,{\boldsymbol n}_a^j(\tilde{t}_n)
 +\Delta\tilde{t}\,\dot{C}_1(\tilde{t}_{n}){\boldsymbol e}_1, 
 ~\text{for}~j=1,\ldots,N_a-1,~\text{and}
 \label{position_update_junction_2D}
\end{equation}
\begin{equation}
 \tilde{\boldsymbol x}_b^{j}(\tilde{t}_{n+1})=\tilde{\boldsymbol x}_b^{j}(\tilde{t}_n)
 +\Delta\tilde{t}\,\tilde{V}_b^j(\tilde{t}_n)\,{\boldsymbol n}_b^j(\tilde{t}_n), 
 ~\text{for}~j=1,\ldots,N_b-1,
 \label{position_update1_junction_2D}
\end{equation}   
where $\Delta\tilde{t}$ is the time step, $\tilde{V}_a^j$ and $\tilde{V}_b^j$ are the normal velocities of 
${\mathscr C}_a$ and ${\mathscr C}_b$ given by 
the non-dimensionalized versions of \eqref{GB_velocity_tricrystala} for $a=1,2$ and
\eqref{kinetic_tricrystal_planar311} for $b=3,4$. Details of the discretization for
$\tilde\kappa_i^j$, ${\boldsymbol n}_i^j$ etc. can be seen from \cite{fischer1}. The Rectangle
rule for integration has been used in \eqref{grain_rotation_tricrystala2}, 
\eqref{grain_rotation_tricrystala42}, and \eqref{kinetic_tricrystal_planar4} to compute the 
non-dimensional rotation rate and the translation rate of grains $G_1$ and $G_2$, respectively. The end point 
velocities $\tilde{V}_i^{0}$ and $\tilde{V}_i^{N_i}$, which are used to evaluate the junction 
angles, are computed following \cite{fischer1}. The position vector 
$\tilde{\boldsymbol x}_{\delta}$ of $J_\delta$ is updated using 
\begin{equation}
 \tilde{\boldsymbol x}_{\delta}(\tilde{t}_{n+1})=\tilde{\boldsymbol x}_{\delta}(\tilde{t}_n)
 +\Delta\tilde{t}\,(\tilde{\boldsymbol q}_{\delta}(\tilde{t}_n)+\dot{C}_1{\boldsymbol e}_1),
 \label{position_update_junctions_2D}
\end{equation}  
where $\tilde{\boldsymbol q}_{\delta}$ is given by the non-dimensional version of 
\eqref{junction_velocity12_stress} when the junction mobility is finite and by
\eqref{junction_velocity_stress_minfty} when the mobility is infinite. 

We now present the simulation results for the tricrystal arrangement. At first, we 
ignore the external stress and study GB capillary driven dynamics. Next we incorporate the effect 
of applied shear stress and compare the results with those obtained without it. We also consider a
bicrystal with an embedded grain having asymmetric cross-section and demonstrate the effect
of external shear stress on coupled GB dynamics. The numerical scheme for such closed GB can 
be obtained from the one described above in a straightforward manner. 
For the tricrystal we choose the initial discretization of ${\mathscr C}_a$ as $N_a=100$ and 
${\mathscr C}_b$ as $N_b=50$ grid points. The embedded grain $G_1$ is 
initially taken to be circular with radius $\tilde{R}_a(0)=0.4$.   The initial orientation of the 
grains are taken as $\psi_1 
=14^\circ$, $\psi_2=0^\circ$, and $\psi_3 = 60^\circ$. The initial misorientations are 
therefore $\theta_1=14^\circ$, $\theta_2=44^\circ$, and $\theta_3=30^\circ$. During the 
coupled motion, only $\psi_1$ (and hence $\theta_1$ and $\theta_2$) is allowed to changed 
while others are kept constant. As a sign convention, if any of the misorientation angles turns 
out to be negative,
we add $90^\circ$ to them to obtain an equivalent angle in the range 
$0\leq \theta_i< 90^\circ$, recalling that the considered crystals posses a four-fold 
symmetry \cite{trautt3}. We discretize the GB in the bicrystal initially with $N=100$ grid points 
and consider the initial misorientation to be $8^\circ$. All the computations are done in a domain of 
size $[-1,1]\times [-1,1]$, with time step $\Delta\tilde{t}$ as $10^{-5}$ and $10^{-4}$ for
the case of GB migration and coupled GB motion, respectively. All the GBs are assumed to be 
$[001]$ tilt boundaries. To avoid mesh points coming very close to each other or 
moving far away after time integration, we re-mesh the GBs after every iteration so as to maintain 
accuracy and stability in all the numerical calculations.

\begin{figure}[t!]
\centering
  \includegraphics[width=4in, height=2in] {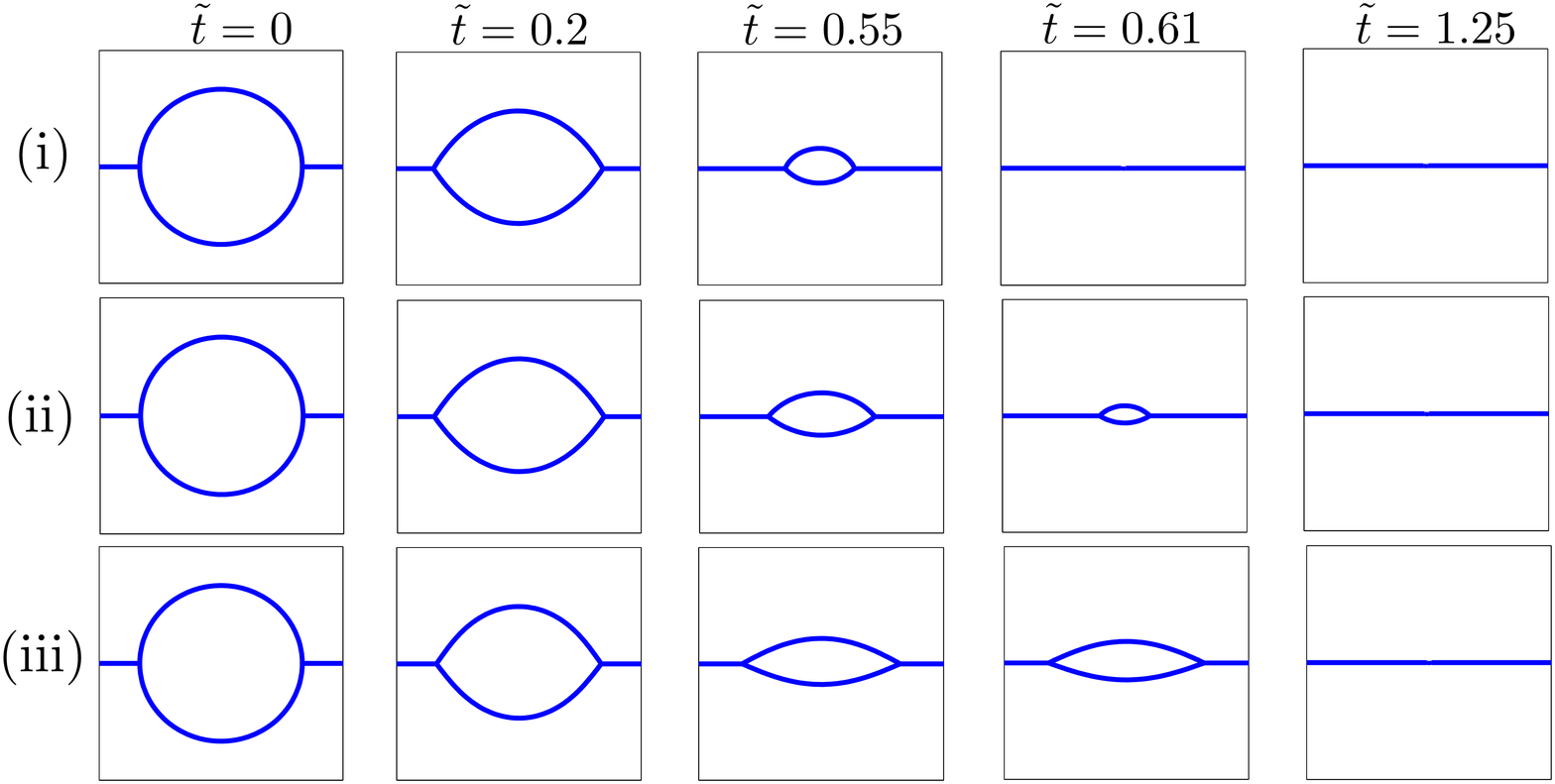}
 \caption{\small{Shape evolution under GB migration when $r_1=0.01$, and 
 $\psi_1 =14^\circ$, $\psi_2=0^\circ$,
 and $\psi_3 = 60^\circ$. Rows (i) to (iii) correspond to $\Lambda\to\infty$, $\Lambda=20$, 
 and $\Lambda=1$, respectively. }}
\label{normal_motion_Lambda_compare}
\end{figure}

\begin{figure}[t!]
\centering
\includegraphics[width=4.1in, height=2.1in] {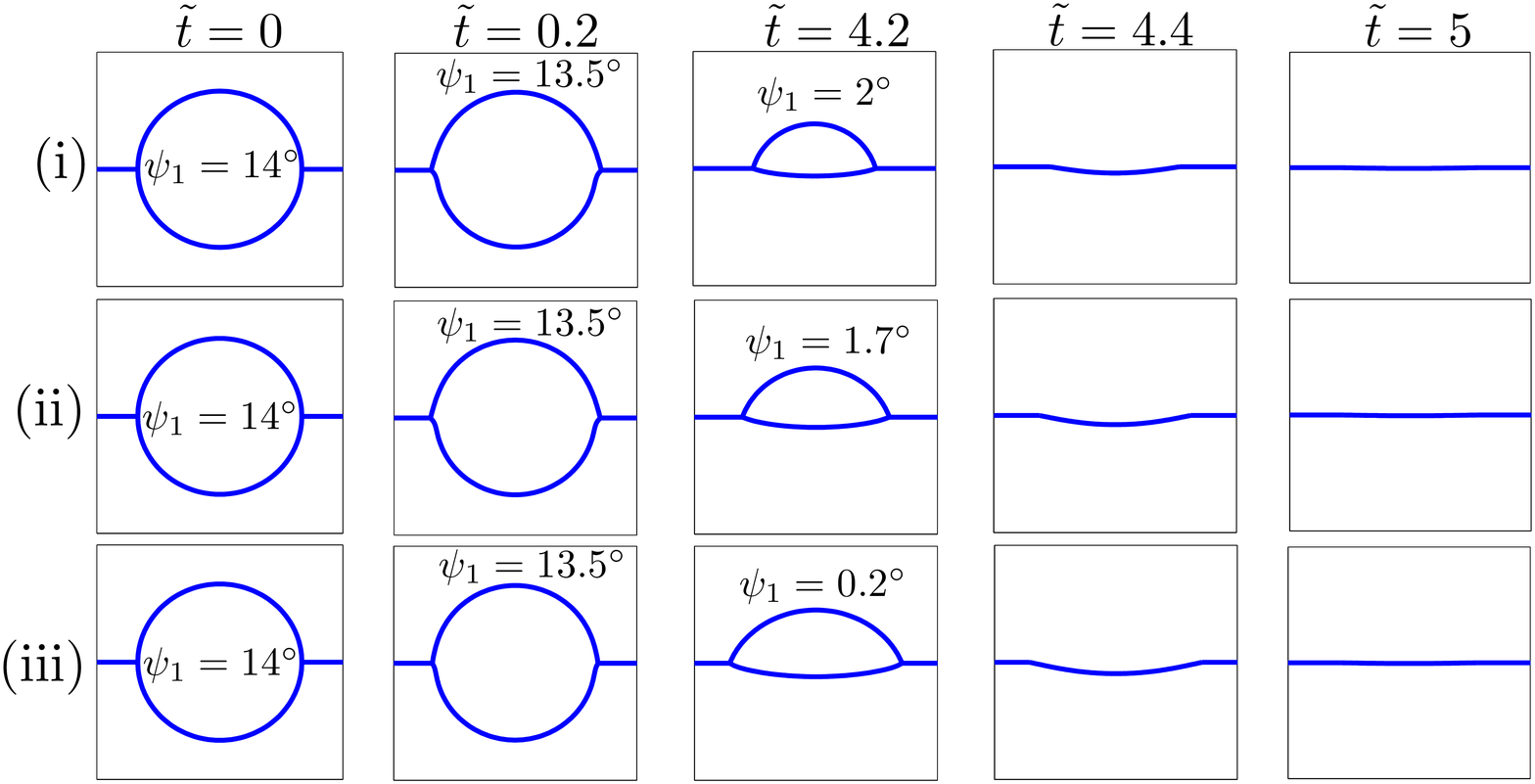}
\caption{\small{Shape evolution under fully coupled GB motion when 
$r_1=0.01$, and initial $\psi_1 =14^\circ$, 
 $\psi_2=0^\circ$ and $\psi_3 = 60^\circ$. Rows (i) to (iii) correspond to $\Lambda\to\infty$,
$\Lambda=20$, and $\Lambda=1$, respectively.}}
\label{coupled_Sbeta_nonzero_Lambda_infty_20_1}
\end{figure}
\subsection{GB Capillary driven motion}
\label{capillary_driven_GBdynamics_res}
We begin by ignoring the applied stress and restrict our attention to the dynamics being driven solely by GB capillary.
The translational velocities of the grains are also neglected.
We present the results only for the tricrystal since bicrystals with an embedded grain have been extensively
studied within the present context \cite{cahn1,taylor1,basak1}.
Note that if ${\mathscr C}_3$ and ${\mathscr C}_4$ are initially planar then they will always remain
stationary, i.e. $V_3=0$ and $V_4=0$, fixing the junction angles $\alpha_3$ and 
$\alpha_4$ for all times. The junction angles ${\alpha}_1$ and $\alpha_2$ and the junction velocities
in such a situation (where atleast one GB at the junction remains stationary) can not be directly calculated using 
\eqref{junction_compatibility_j1}$-$\eqref{junction_velocity_stress_minfty}. The pertinent equations can however
be easily derived, see e.g. Section 3.3 of \cite{fischer1}.

\subsubsection{GB migration} With $\beta \to 0$ and ${S}\to 0$ the kinetic relations \eqref{GB_velocity_tricrystala}, \eqref{kinetic_tricrystal_planar311}, and 
\eqref{grain_rotation_tricrystala2}  are reduced to 
$\tilde{V}_i=\tilde{M}_i\tilde\gamma_i\tilde\kappa_i/2$ and ${\dot \psi}_1=0$, respectively.
Figure \ref{normal_motion_Lambda_compare} shows the evolution of the embedded grain under 
these assumptions with both finite and infinite junction mobility. The junction angles start 
evolving soon after the evolution starts and the embedded grain attains a lens shape. A finite
junction mobility drags the GB motion and retards the shrinking rate of the embedded grain. 
The drag effect increases as $\Lambda$ decreases and the curved GBs become increasingly flatter
before shrinking (see also Figure \ref{area_orientation_embedded_grain}). However, the junction 
velocities become comparable with those of the GBs when $\Lambda>>1$, which reduces the drag on 
the GBs. The area evolution then becomes nearly linear and the deviation from linearity increases 
as $\Lambda$ decreases. The effect of finite junction mobility has been widely noticed to have 
a significant influence on GB dynamics \cite{czubayko1}. The drag effects 
at the junctions are due to frequent dislocation reactions and changes in point defect density 
in their vicinity (Chapter 3 in \cite{koch1}). 

\subsubsection{Coupled GB motion} Depending on the operating conditions, some of the kinetic 
parameters 
may be more active than the others. For example, at temperatures near the melting point, 
viscous
GB sliding dominates over geometric coupling, whereas at relatively lower temperatures, sliding 
is much less active than geometric coupling \cite{cahn2}.  We demonstrate the effect of kinetic 
coefficients on the shape evolution by considering several cases below. 

\medskip

\begin{figure}[t!]
\centering
\subfigure[ ]{
  \includegraphics[width=3.35in, height=1.8in] {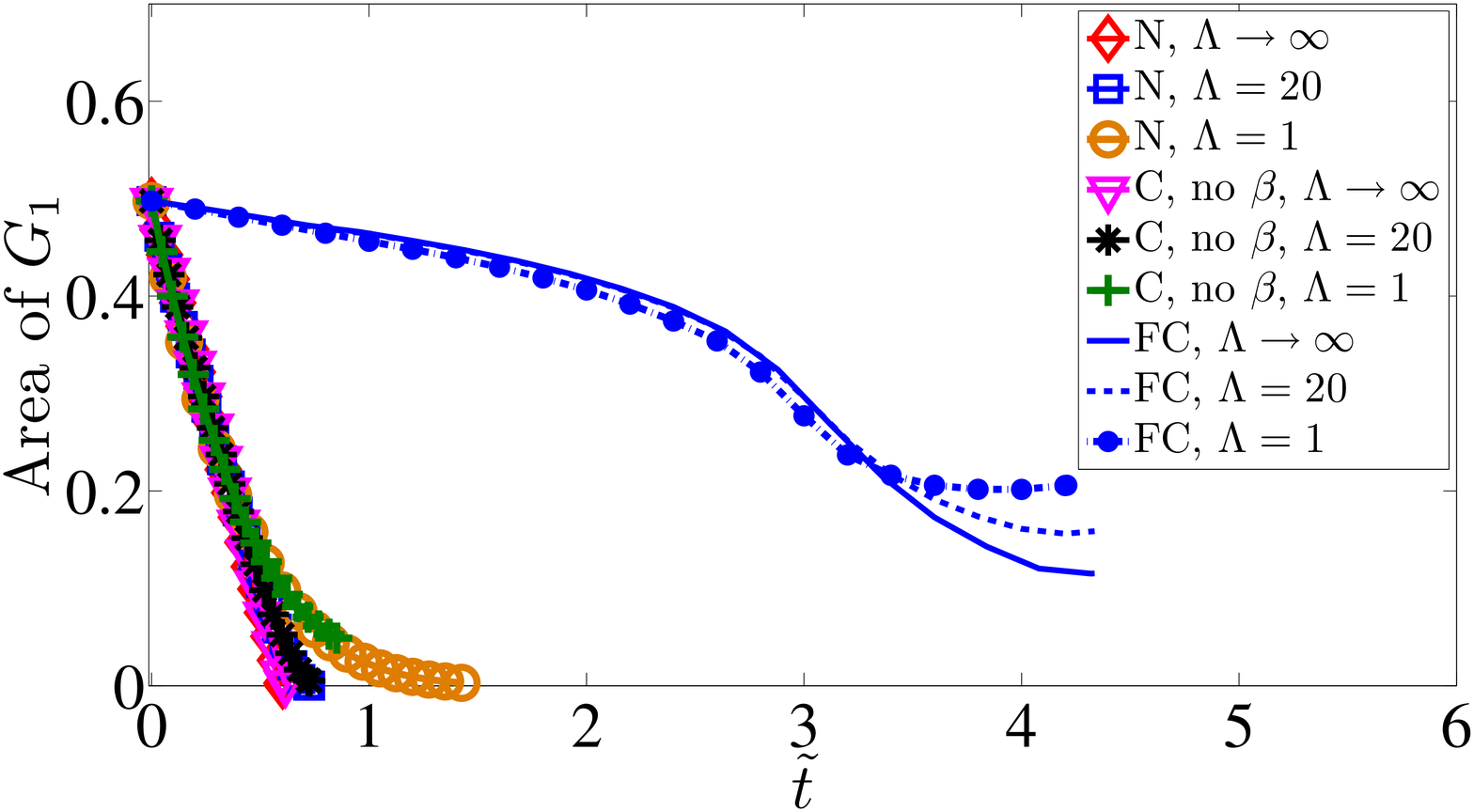}\quad
		\label{area_plot_normal_coupled}}\nonumber
		    \subfigure[ ]{
    \includegraphics[width=3.35in, height=1.85in] {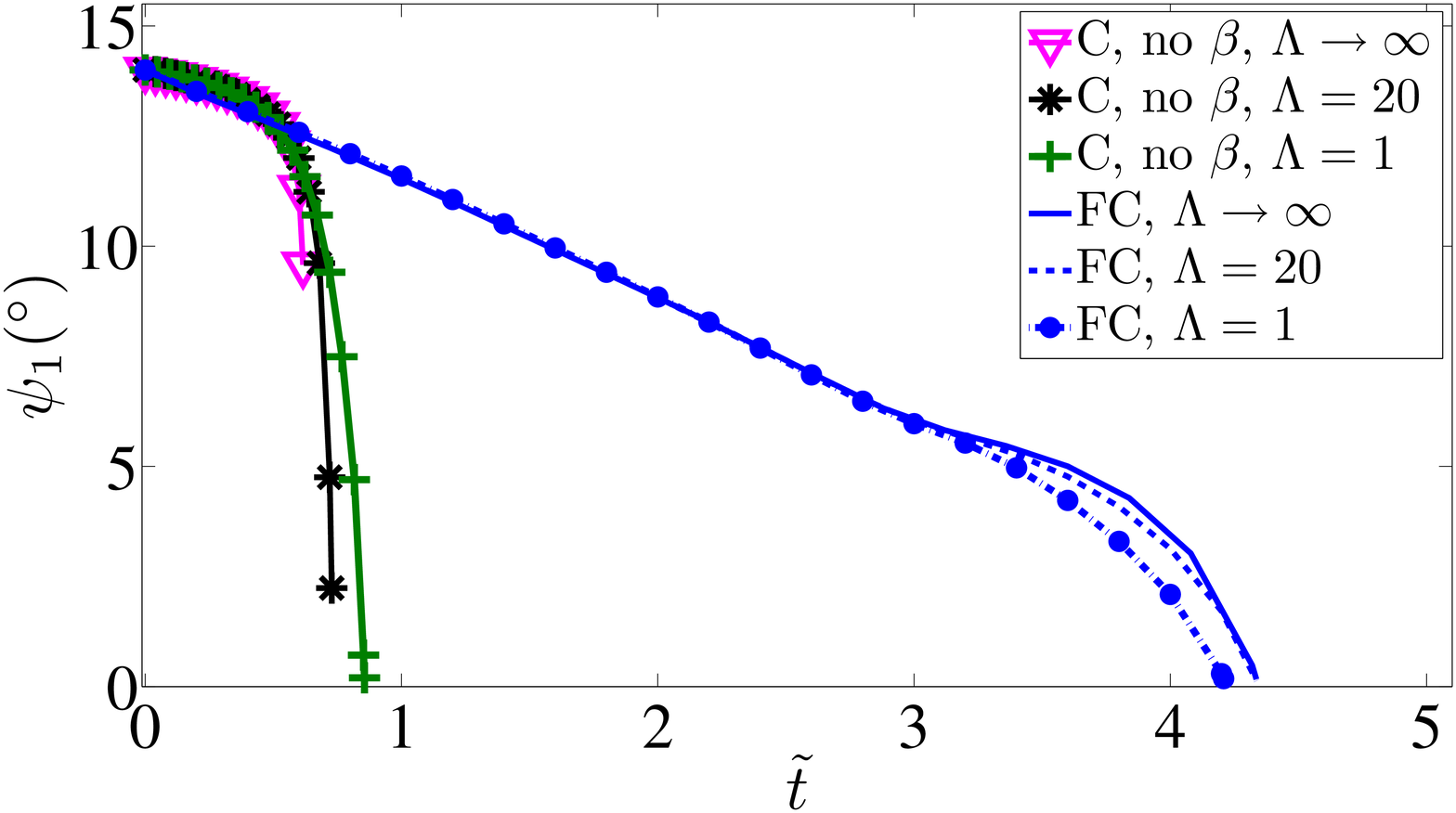}
    \label{orientation_angle1_plot_normal_coupled}}
		\caption{\small{(a) Area and (b) orientation evolution of the embedded grain 
		under normal and 
coupled GB motion when $r_1=0.01$.  Abbreviations: N - normal GB motion, C - coupled GB motion 
in absence of $\beta_1$ and $\beta_2$, and FC - fully coupled GB motion.}}
\label{area_orientation_embedded_grain}
\end{figure}

\noindent \textit{Fully coupled}: When both sliding and geometric coupling are active, the 
grain shrinkage becomes much slower than with GB migration alone, as shown in Figures 
\ref{coupled_Sbeta_nonzero_Lambda_infty_20_1} and \ref{area_orientation_embedded_grain}. 
However, the combined effect of the GB energy and the kinetic coefficients is such that the 
lower GB shrinks faster than the upper one. The dihedral angles between ${\mathscr C}_1$ 
and ${\mathscr C}_2$ are 
greater
in this case than those associated with GB migration at the same time instance  (see Figures 
\ref{normal_motion_Lambda_compare} and \ref{coupled_Sbeta_nonzero_Lambda_infty_20_1}). The 
grain $G_1$ will disappear after it has shrunk to a vanishing volume leaving a bicrystal in 
place of the tricrystal. The embedded grain can also disappear, much before it shrinks to a 
vanishing size, whenever either $\theta_1$
or $\theta_2$ becomes zero; this is in fact the observed situation in Figure 
\ref{coupled_Sbeta_nonzero_Lambda_infty_20_1} and all other considered simulations except 
when the motion is uncoupled. We also note that the finite junction mobility not only drags 
the GB motion, but also slows down the grain rotation, as can be seen in Figure 
\ref{area_orientation_embedded_grain}(b).

\medskip

\noindent \textit{No geometric coupling}: In the absence of $\beta$, the non-dimensional 
equation for 
normal velocity  reduces down to $\tilde{V}_a= \tilde{M}_a\tilde\gamma_a\tilde\kappa_a/2$, 
which is same as the evolution equation for GB migration, except that $\tilde\gamma_a$ 
is now evolving with time (due to evolving misorientation). 
Figure \ref{area_orientation_embedded_grain}(a) shows that the area evolution is now slightly
slower than in the case of GB migration. Orientation $\psi_1$ evolves very slowly for most of the 
time except towards the end.  The shape evolution of the curved GBs is nearly identical to the 
ones shown in Figure \ref{normal_motion_Lambda_compare} for respective junction mobilities. When 
$\Lambda\to\infty$ and $\Lambda=20$, the grain shrinks before $\psi_1$ could vanish. However, when 
$\Lambda=1$, $\psi_1$ vanishes before the area leaving a bicrystal with a depression on the planar 
GB, which also eventually vanishes. 

\medskip

\noindent \textit{No sliding}: For $S\to 0$ \eqref{GB_velocity_tricrystala} implies that the GB 
shape, given by $R_a(\phi_a,t)$, remains self-similar for all times as long as $\beta$ is isotropic 
\cite{taylor1, basak1}.
 For example, if $G_1$ is initially a circle, then it should remain so for all times during the 
 evolution. Obviously with such a restriction, compatibility equations 
 \eqref{junction_compatibility_j1} and \eqref{junction_compatibility_j2} or 
 \eqref{junction_compatibility_j11}, \eqref{junction_compatibility_j12}, and 
 \eqref{junction_compatibility_j13}  will
 have solutions only for very special initial geometries of ${\mathscr C}_1$ and ${\mathscr C}_2$. 

\medskip

\begin{figure}[t!]
\centering
\hspace{-5mm}
\subfigure[]{
  \includegraphics[width=3.35in, height=1.8in] {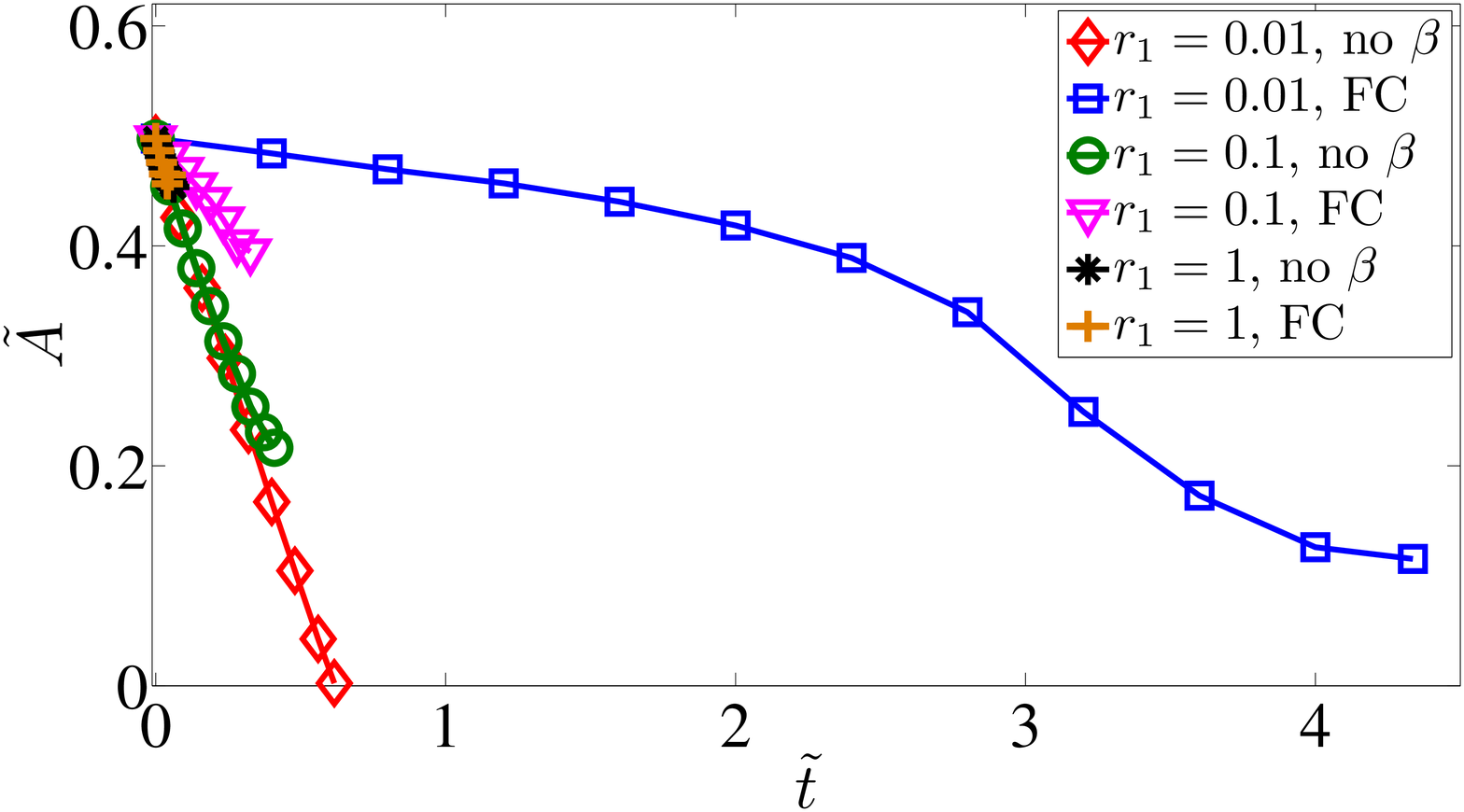}
	\label{area1_varying_r1_junc}}
\hspace{-5mm}
    \subfigure[]{
    \includegraphics[width=3.35in, height=1.8in] {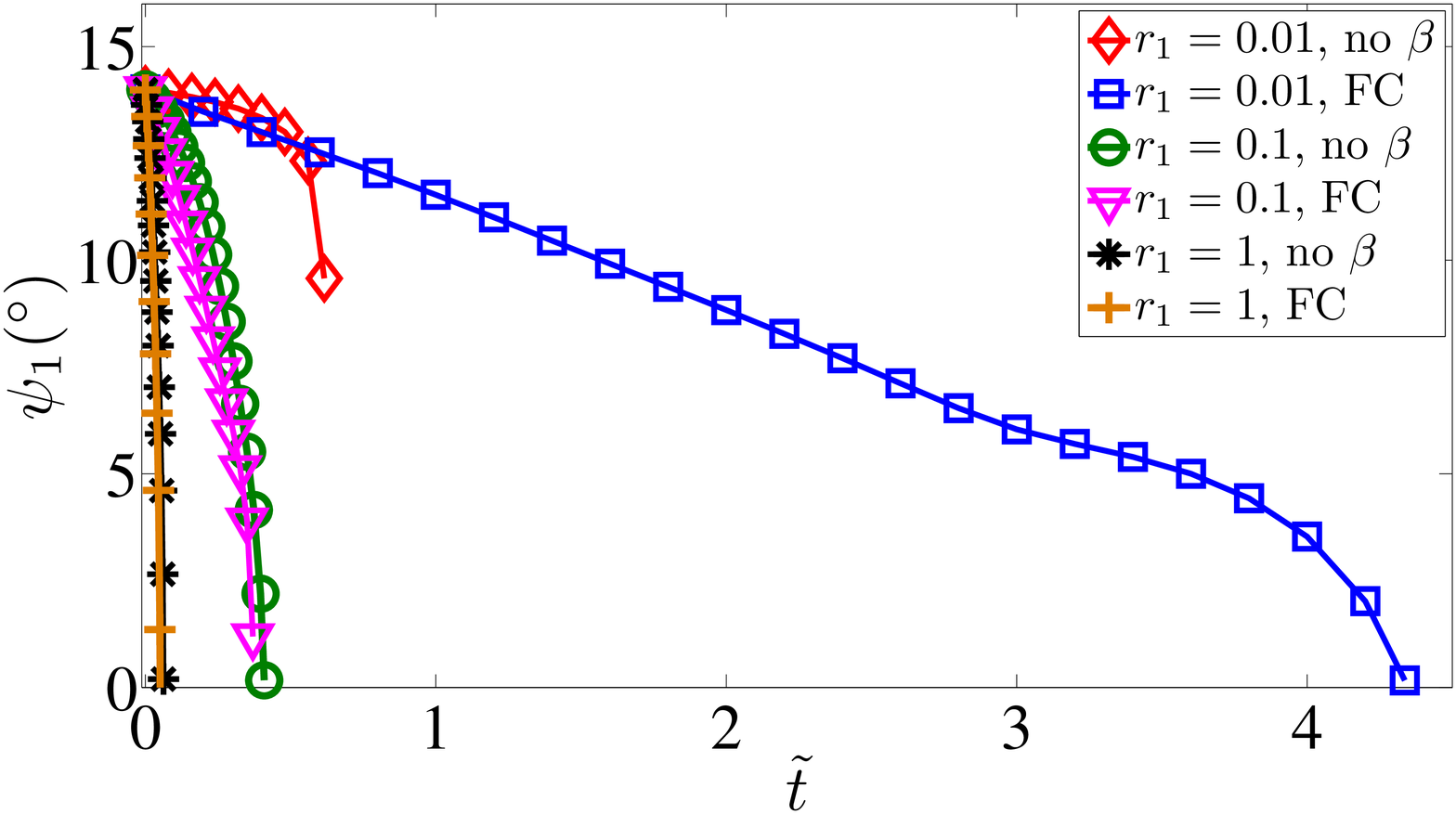}
    \label{psi1_varying_r1_junc}}
\caption{\small{A comparative study of (a) area and (b) orientation evolution of the embedded 
grain under coupled GB motion for varying $r_1$ when $\Lambda \to \infty$.}}
\label{area_psi1_variation_embedded_grain}
\end{figure}

\noindent \textit{Role of sliding}: Higher $r_1$ signifies a relative increase of viscous 
sliding over GB mobility, which is usually seen at elevated temperatures \cite{cahn2}. The 
rate of change of area and orientation $\psi_1$ significantly increases when $r_1$ increases
as shown in Figures \ref{area1_varying_r1_junc} and \ref{psi1_varying_r1_junc}.

 \medskip
 
\noindent \textit{Symmetrically equivalent curved GBs}:
Let us take orientation $\psi_3$ to be $28^\circ$ while keeping initial values of $\psi_1$ and 
$\psi_2$ same as 
above. The initial misorientations are therefore $\theta_1=14^\circ$, $\theta_2 = 76^\circ$, 
and $\theta_3=62^\circ$. The curved GBs are now symmetrically equivalent with 
$\beta_1=-\beta_2$. Since the embedded grain is initially symmetric about ${\boldsymbol e}_1$-axis,
the first term in the numerator of \eqref{grain_rotation_tricrystala2} disappears. However, for 
the GB energy 
considered here, the second term in the numerator will always lead to a non-zero rotation of $G_1$. 
On the 
other hand, if the energy is symmetric about $\theta=45^\circ$ (as is the case with the energy 
given in Figure 6 of \cite{shih1}), the rotation of $G_1$ will vanish and the grain will shrink 
purely by migration of ${\mathscr C}_1$ and ${\mathscr C}_2$. This phenomenon of rotation getting 
locked has been 
observed in the MD \cite{trautt3} and phase field simulations \cite{wu1} when ${\mathscr C}_1$ and 
${\mathscr C}_2$ are symmetrically equivalent. 

\subsection{Effect of stress on GB dynamics}
\begin{figure}[t!]
\centering
\includegraphics[width=4.5in, height=2.3in] {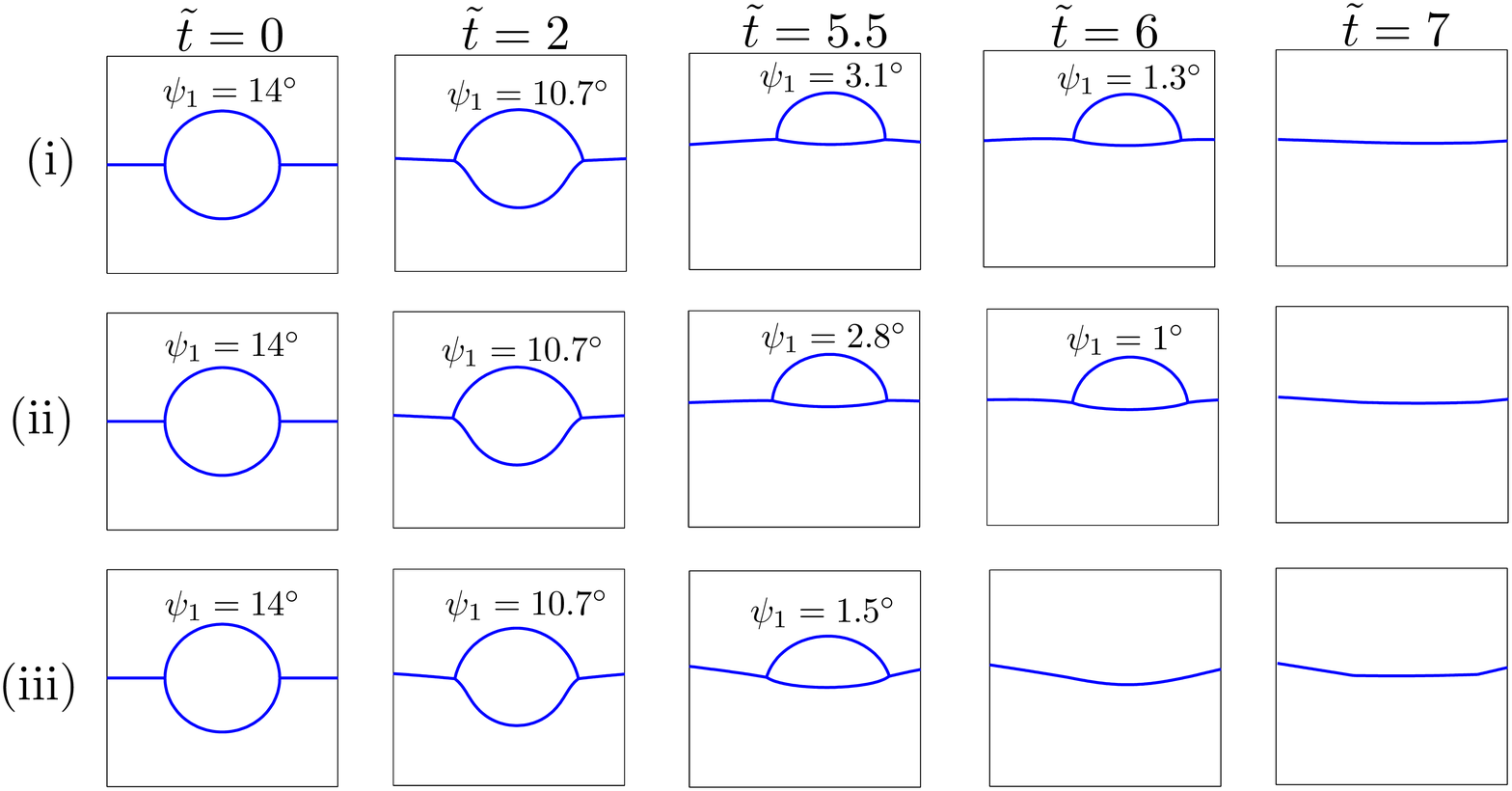}
\caption{\small{Shape evolution and dynamics of the embedded grain under 
fully coupled motion due to the combined effect of GB capillary force and shear
stress of magnitude $\tilde\tau=0.1$ for (i) $\Lambda=100$, (ii) $\Lambda=20$, and
(iii) $\Lambda=1$.}}
\label{shapes_tricrystal_stress}
\end{figure}
\begin{figure}[t!]
\centering
\subfigure[]{
  \includegraphics[width=3.2in, height=1.8in] {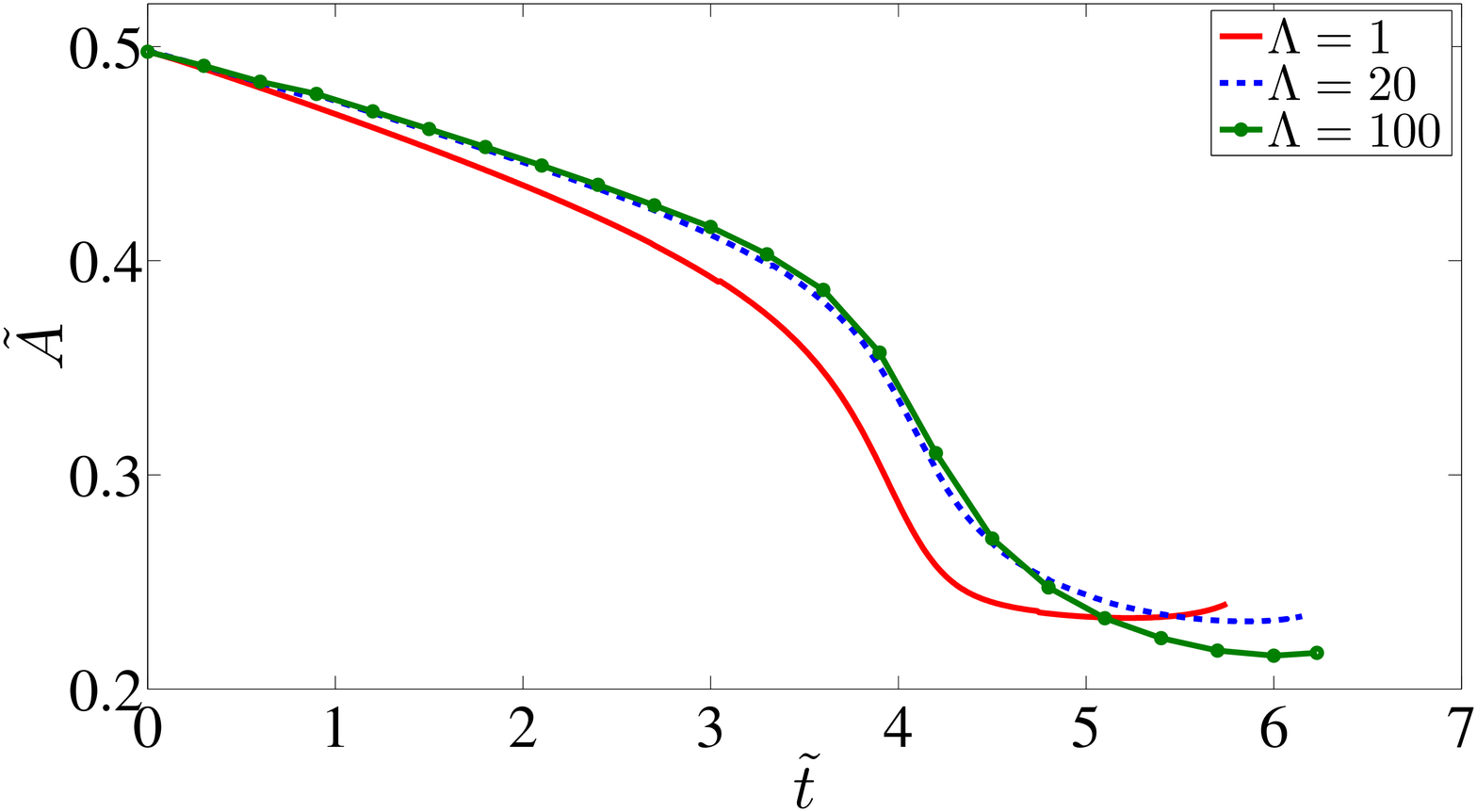}
	\label{area_grain1_stress_vary_Lambda}}
    \subfigure[]{
    \includegraphics[width=3.2in, height=1.8in] {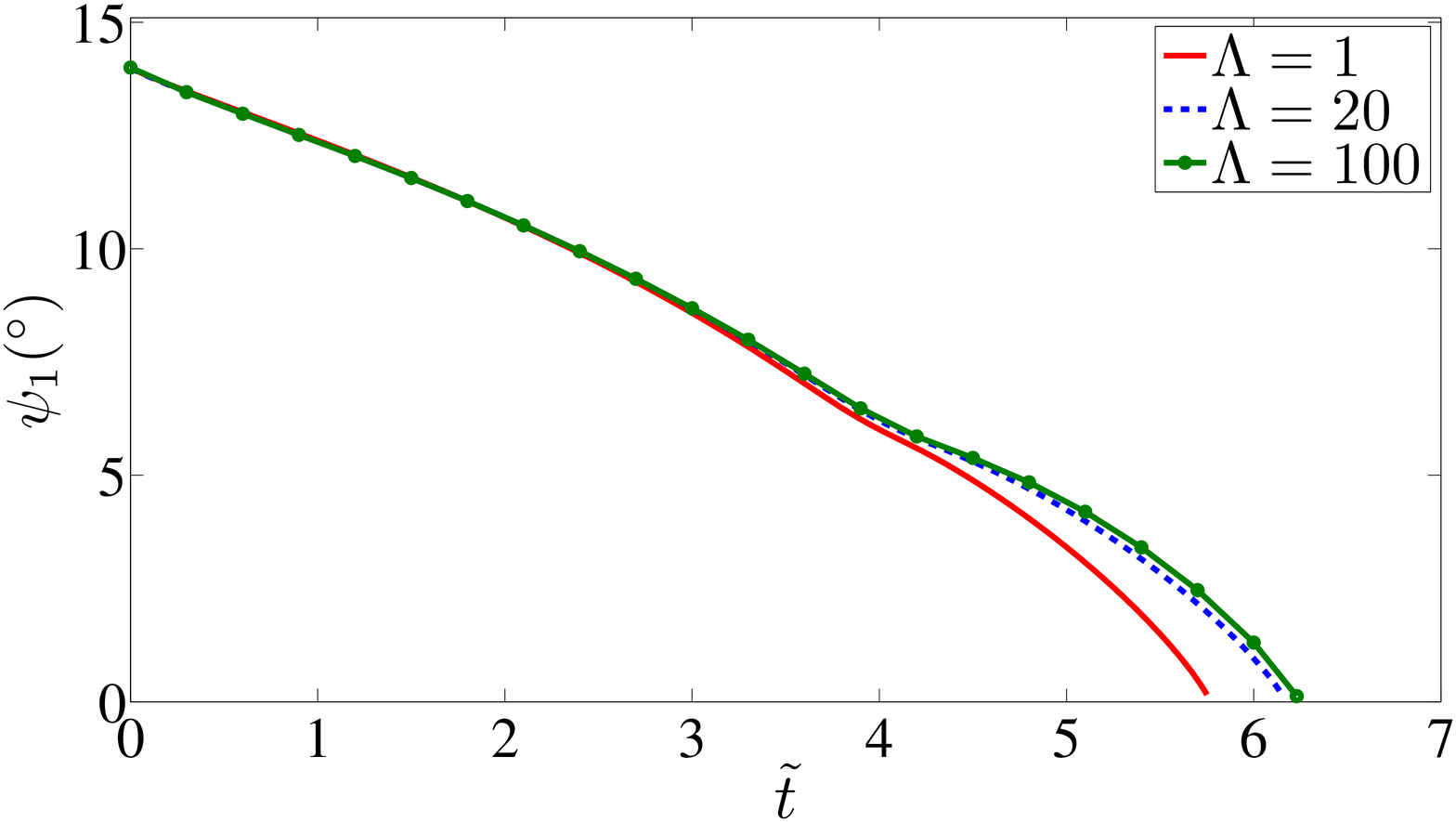}
    \label{orientation_stress_grain1_vary_Lambda}}
\caption{\small{A comparative study of (a) area and (b) orientation evolution of  the embedded 
grain under fully coupled motion (data same as in Figure \ref{shapes_tricrystal_stress}).}}
\label{area_psi1_variation_embedded_grain1}
\end{figure}
Finally, we investigate the effect of shear stress $\tau$ on the coupled GB dynamics in the tricrystal
shown in Figure \ref{tricrystal_2D_stress}. We assume that the dynamics is fully coupled.
We consider $\tilde\tau=0.1$, which implies that $\tau$ is 
approximately equal to $10$ MPa when $\gamma_0=1\,{\rm N/m^2}$ 
and $R_0=10$ nm.  The stress value is much lower
than the yield stress which is of the order of few GPa in NC materials \cite{meyers1}.
All the other kinetic and geometric data have been kept same as considered above in Section 
\ref{capillary_driven_GBdynamics_res}. 
Figure \eqref{shapes_tricrystal_stress} shows the evolution of GB, grain, and junction dynamics
in the tricrystal. The overall evolution is now slower as compared to what was observed during purely
GB capillary driven dynamics in Section \ref{capillary_driven_GBdynamics_res}. The center of 
rotation of
the embedded grain can also be seen to translate from the initial position. With increasing junction 
mobility the magnitude of translation increases. 
Figures \ref{area_grain1_stress_vary_Lambda} and \ref{orientation_stress_grain1_vary_Lambda} 
show that decreasing junction mobility marginally increases the rate of grain rotation and 
grain shrinkage, which is in fact opposite to what has been observed when the dynamics is driven only 
by GB capillary. This can be attributed to
the additional effects coming from the stress related term in \eqref{grain_rotation_tricrystala2}
and the non-trivial curvature generated in ${\mathscr C}_3$ and ${\mathscr C}_4$ due to large 
junction drag when $m_\delta$ is small. All the cases considered in Figure 
\ref{shapes_tricrystal_stress} show that vanishing of the misorientation at ${\mathscr C}_1$, due 
to the rotation of the embedded grain, leaves behind a depression on the GB separating
the rectangular grains. The depression ultimately disappears so as to eliminate the curvature.

\begin{figure}[t!]
\centering
\includegraphics[width=4.1in, height=2.1in] {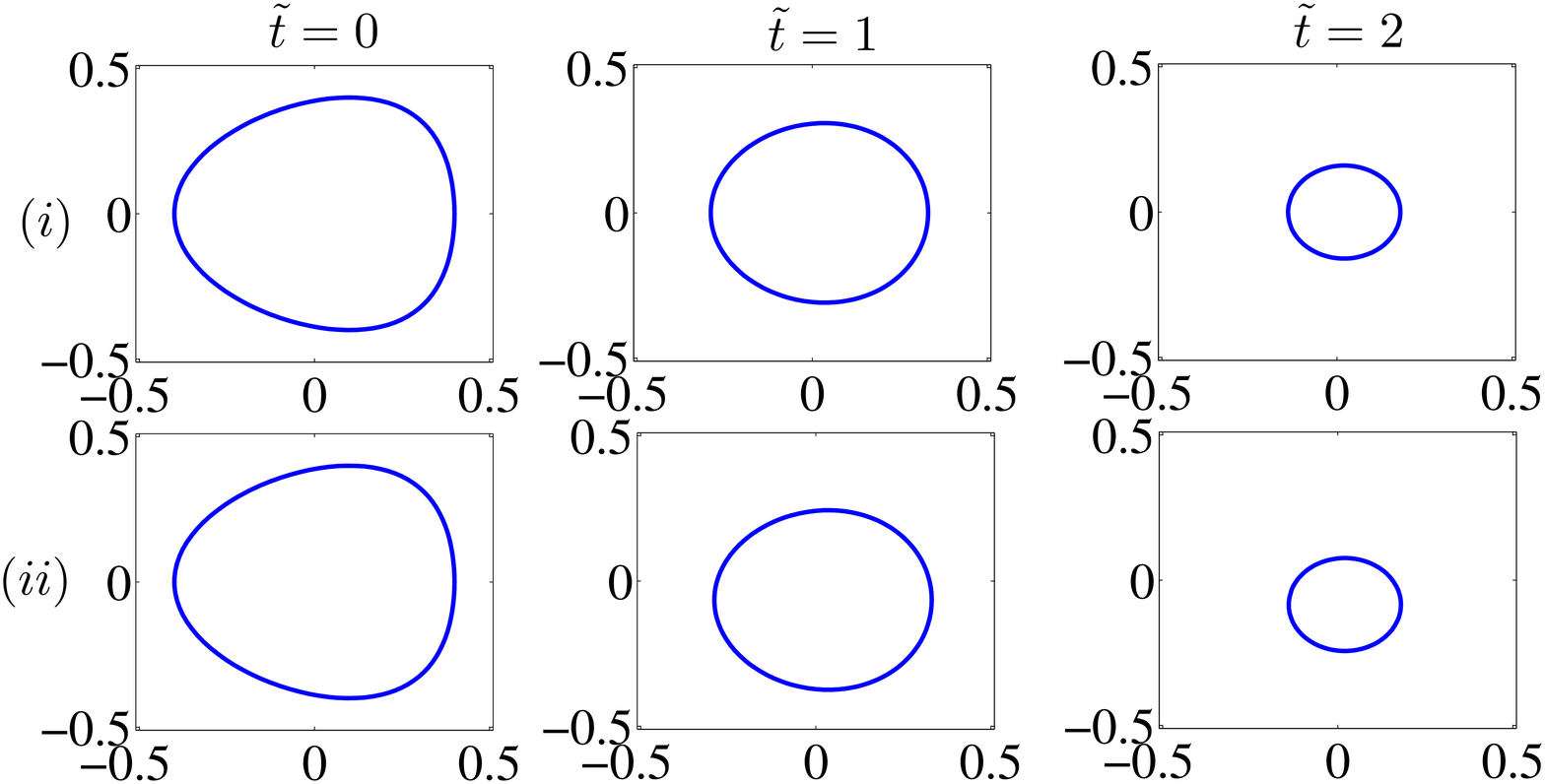}
\caption{\small{Shape evolution and dynamics of the embedded grain in a bicrystal (i) when 
the external stress is absent, and (ii) when the shear stress is $\tilde\tau=30$. We take $r_3=50$.
All the other parameters are same as in Section \ref{capillary_driven_GBdynamics_res}.}}
\label{shape_evol_bicrystalII.eps}
\end{figure}

We end our study by noting the effect of applied shear stress $\tau$ on the shape evolution of 
a bicrystal arrangement as described in Remark 2 at the end of previous section. We considered the geometry of ${\mathscr C}$ as shown at 
$\tilde{t}=0$ in
Figure \ref{shape_evol_bicrystalII.eps}. The magnitude of the stress $\tilde\tau=0.1$,
considered previously for the tricrystal arrangement, does not
make any significant difference to the GB dynamics when compared to that observed in the 
absence of stress. As a result we assume a higher stress $\tilde\tau=30$ (i.e. $\tau=3$ GPa). Figure \ref{shape_evol_bicrystalII.eps} shows comparison of the shape evolution 
of the embedded grain when applied stress is absent and when the bicrystal is subjected to 
shear stress. Clearly the center of rotation of the embedded grain in the latter
case is translating, whereas in the former it is fixed.

\section{Conclusions}
\label{conclusion_junction_2D}
We have extended the analytical study of coupled GB motion, hitherto 
restricted to bicrystals with a columnar grain having a fixed center of rotation embedded 
in a larger grain, 
by introducing triple junctions and relative translational sliding in the analysis. The 
present formulation 
is applied to a tricrystal (and a bicrystal) without restricting the center of rotation of the 
embedded grain 
to be fixed. In deriving the necessary kinetic relations we have provided a novel 
thermodynamic framework within which such and more complicated incoherent interfaces can be 
studied. Our thermodynamic formalism is closely related to earlier work on incoherent 
interfaces, most notably \cite{cermelli2, gupta1}. The present work can be extended in several
directions: i) to analyze the coupled motion in three dimensions, ii) to include grain 
deformation in terms of 
elastic/plastic behaviour of the grains, iii) to include bulk diffusion. While we have 
considered a simpler case in 
two-dimensions ignoring these effects, we are still confronted with a formidable boundary 
value problem which can be solved satisfactorily only under some further assumptions. For 
instance when considering anisotropic GB energies a more sophisticated numerical technique 
(such as the level set method) is needed \cite{basak1}. However, including junction dynamics 
within a level set framework remains unsolved except for some very specific cases, restricted
to constant interfacial energy and kinetic coefficients along with infinite junction mobility.
This led us to consider only isotropic energies so that the resultant problem with junctions 
is solvable through a simpler numerical scheme. We should also point out that the linear 
kinetic relations developed in this work are capable of capturing the physical 
phenomenon only close to the equilibrium. Our aim is to present a rigorous framework for 
dealing problems of great utility in polycrystalline materials and to demonstrate the efficacy
of the proposed set of governing equations using simple bicrystal and tricrystal arrangements,
motivated by recent MD simulation studies. Although the 
tricrystal system is much simpler than the real polycrystal which would consist of 
numerous grains (generally polyhedral) and triple junctions, we expect 
the essential features of the model, like drag induced by junctions on GB motion and grain 
rotation, to remain valid. In any case, extending the present formulation to a real 
polycrystal with many grains is only a problem of greater computational effort and should 
be straightforward.

\section*{Acknowledgement}
The authors gratefully acknowledge Jiri Svoboda for helpful discussions.

\bibliography{stress_junction}
\bibliographystyle{plain}
 \appendix
 \section{Balance laws}
 \label{balance_laws_stress_2D1}
 In this appendix we derive the mass balance and linear momentum balance relations for grains,
 GBs, and junctions, all of which are used in deriving the local dissipation inequalities in Section 
 \ref{thermo_junction_2D}.  We assume that the bulk field $f$ defined in Section \ref{thermo_junction_2D} 
 satisfies the  following limit (see Appendix A of \cite{simha1}):
\begin{equation}
\int_{P} f\,da = \lim_{\epsilon\to 0}\int_{{P}_\epsilon}f\,da,
 \label{regularity_field_junc_2D}
\end{equation}
where $da$ is an infinitesimal area element from the region $P$. Using the standard transport 
relations for the bulk 
quantities we can show (cf. Appendix A of \cite{simha1} and Chapter 32 in \cite{gurtin_cont_mech})
\begin{equation}
\frac{d}{dt}\int_{P}f \,da = \int_{P}(\dot{f}+f\nabla\cdot{\boldsymbol v})\,da-\sum_{i=1}^3
 \int_{\Gamma_i}[\![f U_i]\!]\, dl-\lim_{\epsilon\to 0}\int_{{\mathscr C}_\epsilon
 }f\,({\boldsymbol u}-{\boldsymbol v})\cdot{\boldsymbol m}\,dl,
\label{bulk_transport2_2D}
\end{equation}
where the overdot denotes the material time derivative of $f$, $\nabla$ is the gradient operator, 
${\boldsymbol v}$ is the particle velocity, $V_i$ is the normal velocity of $\Gamma_i$, 
$U_i=V_i-{\boldsymbol v}\cdot{\boldsymbol n}_i$ is the relative normal velocity of the GB, 
${\boldsymbol m}$ is the outward normal to the disc ${\mathscr D}_\epsilon$,
and $dl$ is an infinitesimal line element. The term $\nabla\cdot{\boldsymbol v}$ denotes the 
divergence of the velocity field.

 \paragraph{Mass balance} The rate of change of total mass in $P$ in the absence of any 
 external source of mass 
generation/accretion, with a vanishing mass flux across the boundary $\partial P$, should be 
balanced by the mass flux through the edges $A_i$, i.e.
\begin{equation}
\frac{d}{{d}t}\int_P \rho\,da =  -\sum_{i=1}^3(h_i)_{A_i},
\label{bomint_junc}
\end{equation}
where $\rho$ is the mass density of the bulk and $h_i$ is the diffusional flux along $\Gamma_i$ 
in the direction of increasing arc-length parameter $s_i$. 
Using \eqref{bulk_transport2_2D} and the divergence 
theorem (cf. (32.27)$_2$ in \cite{gurtin_cont_mech}) the following 
local mass balance relations are imminent:
\begin{equation}
\dot\rho + \rho\,\nabla\cdot{\boldsymbol v}=0 ~{\rm in}~P_i,
\label{bom_junc}
\end{equation}
\begin{equation}
[\![\rho U_i ]\!]=\frac{\partial h_i}{\partial s_i} ~{\rm on}~\Gamma_i,~\text{and}
\label{boms_junc}
\end{equation}
\begin{equation}
\sum_{i=1}^3 h_i-\lim_{\epsilon\to 0}\oint_{{\mathscr C}_\epsilon}\rho\,({\boldsymbol u}
-{\boldsymbol v})\cdot{\boldsymbol m}\,dl=0 ~{\rm at}~J.
\label{bomjun_junc}
\end{equation}

\paragraph{Linear momentum balance}
Neglecting inertia and body forces, the 
balance of linear momentum requires $\int_{\partial P}{\boldsymbol\sigma}{\boldsymbol m}\,dl=0$,
where ${\boldsymbol\sigma}$ is the symmetric Cauchy stress tensor.
Applying Equation (A5) of \cite{simha1} this global balance law can be reduced to the following 
local equations:
\begin{equation}
\nabla\cdot{\boldsymbol\sigma} = {\boldsymbol 0} ~\text{in}~ P_i,
\label{blm_grain_junc_stress}
\end{equation}
\begin{equation}
[\![{\boldsymbol\sigma}]\!]{\boldsymbol n}_i = {\boldsymbol 0} ~\text{on}~
\Gamma_i,~\text{and}
\label{blms_junc_stress}
\end{equation}
\begin{equation}
\lim_{\epsilon\to 0}\oint_{{\mathscr C}_\epsilon}{\boldsymbol \sigma}{\boldsymbol m} dl
={\boldsymbol 0}~\text{at}~J.
\label{blmjun_junc_stress}
\end{equation} 
According to 
\eqref{blms_junc_stress} the traction is continuous across $\Gamma_i$. On the other hand,
\eqref{blmjun_junc_stress} implies that even with a singular stress at the junction, the net
force on the periphery of ${\mathscr C}_\epsilon$ ($\epsilon\to 0$) is finite. This is known as the 
standard weak singularity condition which requires  ${\boldsymbol\sigma}\sim\epsilon^{-\zeta}$, 
where $\zeta< 1$ (see Chapter 34 in \cite{gurtin_config} for further discussion).

\end{document}